\documentclass[
 reprint,
superscriptaddress,
preprintnumbers,
floatfix,
 amsmath,amssymb,
 aps,
]{revtex4-2}

\usepackage{siunitx}
\usepackage{lipsum}
\usepackage{graphicx}% Include figure files
\usepackage{subfigure}
\usepackage{dcolumn}% Align table columns on decimal point
\usepackage{bm}% bold math
\usepackage{hyperref}
\usepackage{xcolor}
\hypersetup{
    colorlinks = true,
    linkcolor=red,
    citecolor = blue %Change to black if we want them invisible
}

\begin{document}

\preprint{FERMILAB-PUB-24-0025-T}

\title{Dark counts in optical superconducting transition-edge sensors for rare-event searches}% Force line breaks with \\

\author{Laura Manenti}
\email{laura.manenti@nyu.edu}
\affiliation{Division of Science, New York University Abu Dhabi, United Arab Emirates}
\affiliation{Center for Astrophysics and Space Science (CASS), New York University Abu Dhabi, United Arab Emirates}

\author{Carlo Pepe}
\affiliation{
 Dipartimento di Elettronica Telecomunicazioni Politecnico
di Torino, Corso Duca degli Abruzzi 24, 10129 Torino, Italy}
\affiliation{INRiM, Istituto Nazionale di Ricerca Metrologica, Strada delle Cacce, 91 – 10135 Torino, Italy}
\affiliation{INFN Sezione di Torino - Torino, Italy}

\author{Isaac Sarnoff}
\email{sarnoff@nyu.edu} 
\affiliation{Division of Science, New York University Abu Dhabi, United Arab Emirates}
\affiliation{Center for Astrophysics and Space Science (CASS), New York University Abu Dhabi, United Arab Emirates}

\author{Tengiz Ibrayev}
\affiliation{Division of Science, New York University Abu Dhabi, United Arab Emirates}
\affiliation{Center for Astrophysics and Space Science (CASS), New York University Abu Dhabi, United Arab Emirates}

\author{Panagiotis Oikonomou}
\affiliation{Division of Science, New York University Abu Dhabi, United Arab Emirates}
\affiliation{Center for Astrophysics and Space Science (CASS), New York University Abu Dhabi, United Arab Emirates}

\author{Artem Knyazev}
\affiliation{Division of Science, New York University Abu Dhabi, United Arab Emirates}
\affiliation{Center for Astrophysics and Space Science (CASS), New York University Abu Dhabi, United Arab Emirates}

\author{Eugenio Monticone}
\affiliation{INRiM, Istituto Nazionale di Ricerca Metrologica, Strada delle Cacce, 91 – 10135 Torino, Italy}

\author{Hobey Garrone}
\affiliation{
 Dipartimento di Elettronica Telecomunicazioni Politecnico
di Torino, Corso Duca degli Abruzzi 24, 10129 Torino, Italy}
\affiliation{INRiM, Istituto Nazionale di Ricerca Metrologica, Strada delle Cacce, 91 – 10135 Torino, Italy}

\author{Fiona Alder}
\affiliation{Dept. of Physics and Astronomy, University College London, Gower Street, London, United Kingdom}

\author{Osama Fawwaz}
\affiliation{Division of Science, New York University Abu Dhabi, United Arab Emirates}
\affiliation{Center for Astrophysics and Space Science (CASS), New York University Abu Dhabi, United Arab Emirates}

\author{Alexander J. Millar}
\affiliation{Theoretical Physics Division, Fermi National Accelerator Laboratory, Batavia, IL 60510, USA}
\affiliation{Fermi National Accelerator Laboratory, Batavia, IL 60510, USA}

\author{Knut Dundas Morå}
\affiliation{
 Physics Department, Columbia University, New York, New York 10027, USA}

\author{Hamad Shams}
\affiliation{Division of Science, New York University Abu Dhabi, United Arab Emirates}
\affiliation{Center for Astrophysics and Space Science (CASS), New York University Abu Dhabi, United Arab Emirates}

\author{Francesco Arneodo}
\affiliation{Division of Science, New York University Abu Dhabi, United Arab Emirates}
\affiliation{Center for Astrophysics and Space Science (CASS), New York University Abu Dhabi, United Arab Emirates}

\author{Mauro Rajteri}
\affiliation{INRiM, Istituto Nazionale di Ricerca Metrologica, Strada delle Cacce, 91 – 10135 Torino, Italy}

\date{\today}

\begin{abstract}
Superconducting transition-edge sensors (TESs) are a type of quantum sensor known for its high single-photon detection efficiency and low background. This makes them ideal for particle physics experiments searching for rare events. In this work, we present a comprehensive characterization of the background in optical TESs, distinguishing three types of events: electrical-noise, high-energy, and photonlike events. We introduce computational methods to automate the classification of events. For the first time, we experimentally verify and simulate the source of the high-energy events. 
We also isolate the photonlike events, the expected signal in dielectric haloscopes searching for dark matter dark photons, and achieve a record-low photonlike dark-count rate of \num{3.6d-4}\,\unit{\hertz} in the \numrange{0.8}{3.2}\,eV energy range.
\end{abstract}

\maketitle

%%%%%%%%%%%%%%%%%%%%%%%%%%%%%%%
\section{Introduction}
%%%%%%%%%%%%%%%%%%%%%%%%%%%%%%% 
Superconducting transition-edge sensors (TESs) are highly sensitive microcalorimeters, capable of detecting single- and multi-photon events with an efficiency close to unity~\cite{irwin1995application}. Most TESs operate below
$\mathcal{O}(\SI{1}{\kelvin})$, at the transition point between their superconducting and normal states, and can be tailored to different wavelength ranges, spanning from x-ray to infrared.

Several particle-physics experiments are employing---or plan to employ---TESs in rare event searches in the $\mathcal{O}$(0.1--10)\,eV range because of their low background and high detection efficiency. TESs sensitive to photons in this energy range are referred to as optical TESs. Examples of experiments using optical TESs include PTOLEMY~\cite{Cocco2007Jun, PTOLEMYCollaboration2022Sep}, targeting relic neutrinos, and ALPS II, focusing on new light bosons and other weakly interacting subelectronvolt particles~\cite{gimeno2023tes}. 
Understanding the background of TESs is essential for this kind of rare-event searches.
A previous study~\cite{dreyling2015characterization} investigated the background, also referred to as dark-count rate (DCR), for an infrared TES optimized for a wavelength of \SI{1064}{nm}. 
While the authors identified various kinds of events contributing to the intrinsic DCR (counts detected by the TES without any input optics), they did not investigate the nature of these specific signals. They did, however, explore the DCR of a fiber-coupled TES, where the ``warm'' fiber end was located outside the cryostat (see Section 5.2 of Ref.~\cite{dreyling2015characterization}). It is important to note that this latter case represents the total dark-count rate of the device as defined by Ref.~\cite{dictionary}, which differs from the intrinsic DCR that is the focus of our current study.

In our work, we investigate the nature of the signals when no optical fiber is connected to the TES. Moreover, we measure the DCR over a broader wavelength range, spanning from 390 to \SI{1550}{nm}. To assess the nature of the observed background events, we have devised distinct experimental layouts, employing radioactive sources and a cosmic ray coincidence setup.

TESs would also be the ideal photosensors in dielectric haloscope detectors for the detection of dark matter (DM) in the form of dark photons (DPs)~\cite{millar2017dielectric}. Two experiments have recently pioneered the use of dielectric haloscopes to search for DM DPs: LAMPOST~\cite{chiles2022new}, in the USA, and MuDHI~\cite{manenti2022search}, in the United Arab Emirates. The LAMPOST experiment used a superconducting nanowire single-photon detector (SNSPD) as the photosensor, while MuDHI employed a single-photon avalanche diode (SPAD). 
The TES in the present study was designed for use in the future upgrade of the MuDHI experiment, a multilayer dielectric haloscope targeting DPs with mass around \SI{1.5}{eV/c^2}. 
We have measured the DCR of such an optical TES and developed a set of experimental and data-processing methods to characterize, reduce, and mitigate its background rate. Using these methods, we have obtained a DCR of \num{6d-5}\,\unit{\hertz} at $1.5\pm 0.2$\,eV. 
This represents an improvement of approximately 7 orders of magnitude compared to the SPAD used in the previous MuDHI setup. Furthermore, we have identified and explained the primary contributors to optical TES dark-counts, linking them to cosmic and environmental gamma-ray interactions with the TES substrate.

%**********************************************
\section{Superconducting \\transition-edge sensors}
%**********************************************
\subsection{Design and operating principle}

A TES is a thin superconducting film operated at the transition between its superconducting and normal state where it is most sensitive to changes in temperature. The TES working point is stabilized on the transition by negative electrothermal feedback~\cite{irwin1995application} which is achieved by applying a voltage bias. This means that any increase in temperature due to energy absorption leads to an increase in resistance, which in turn causes a decrease in current. This drop in current reduces Joule heating, thereby cooling the sensor and bringing it back to the working point.
A direct current superconducting quantum interference device (dc SQUID) connected to the TES acts as a transimpedance amplifier~\cite{martinis1985signal}. It converts the current drop resulting from the particle interaction to a measurable voltage drop. This voltage can then be amplified by standard electronic amplifiers and the waveforms can be measured and recorded by an oscilloscope. The voltage drop produced by the dc SQUID is proportional to the energy deposition in the TES within its operating range~\cite{irwin1995application}.

The two superconducting TESs used in this work were fabricated and operated at the Istituto Nazionale di Ricerca Metrologica (INRiM) in Turin (Italy). They are made of a bilayer of titanium, the superconducting material, and gold, a normal metal to lower the critical temperature, $T_{\rm c}$, of the titanium (approximately 400 mK) by the proximity effect~\cite{rajteri2020tes}. Both TESs exhibit a $T_{\rm c}$ of around \SI{90}{\milli K} and are operated inside an adiabatic demagnetization refrigerator (ADR). 
Since TESs and dc SQUIDs are sensitive to magnetic fields, including the geomagnetic field, a cryogenic magnetic shield is installed within the ADR around the working area to suppress any magnetic interference that could degrade the performance of the sensor. The magnetic shield is a cylindrical case made of a high-permeability nickel-iron alloy, manufactured by Amuneal, providing shielding capabilities~\cite{amumetal}.

Two 20$\times$\SI{20}{\micro m^2} TESs were deposited on silicon substrates. We refer to the sensors as ``TES~E'' and ``TES~B.'' Most of the data were acquired from TES~E, making it the primary focus of our study. Unless otherwise specified, references to TES data should be assumed to pertain to TES~E. In Appendices~\ref{appendix_TES_fabrication} and \ref{appendix_TESB}, we provide details on the TES fabrication and some information about the characterization and dark-count rate of TES~B.
For both TES~E and TES~B, the bath temperature was consistently maintained at \SI{43}{mK}, with the working point set at 15\% of their normal resistance.
The oscilloscope used for the data acquisition was a LeCroy Wave Runner. The TES signal has only been amplified by a single-stage 16-SQUID series array~\cite{drung2007highly}, using the XXF-1 Magnicon electronics~\cite{Magnicon}.

%**********************************************
\subsection{Calibration}
\begin{figure}[t]
    \centering
    \includegraphics[width=\linewidth]{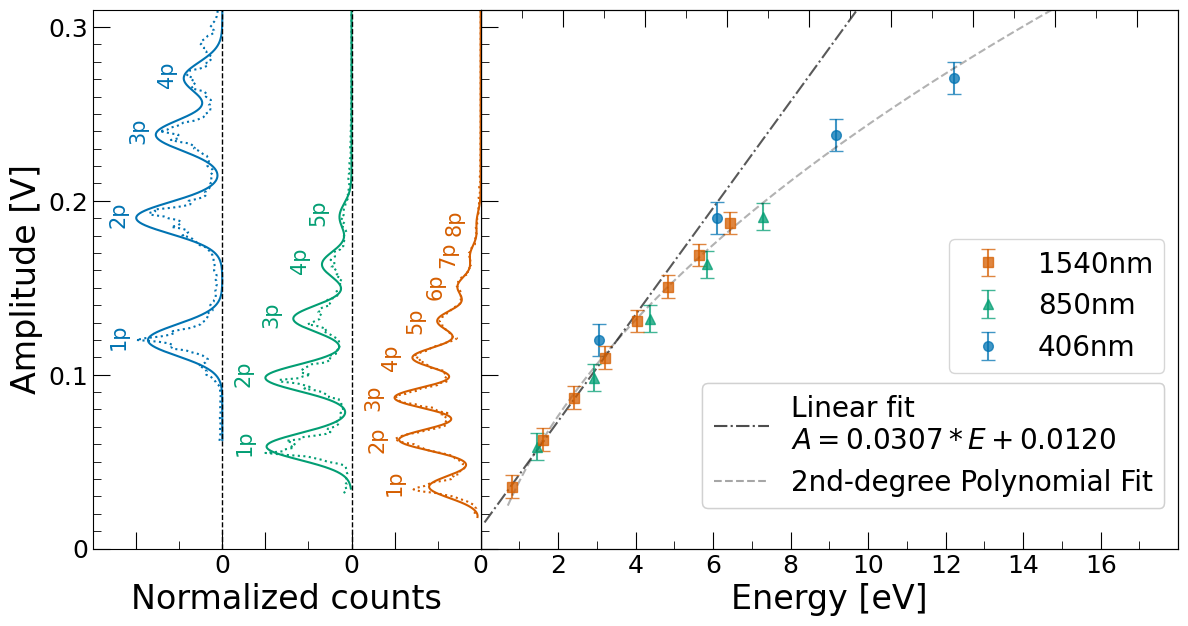}
    \caption[]{Calibration of TES~E using optically coupled, attenuated laser diodes. 
    Right: The calibration curve, with signal amplitude (V) plotted against photon energy (eV). Data points are color-coded by laser wavelength: blue for \SI{406}{nm}, green for \SI{850}{nm}, and orange for \SI{1540}{nm}. Left: Histograms showing count distribution versus amplitude for each wavelength. The total number of counts in each histogram is normalized to 100. A multi-Gaussian fit (solid line) is shown on top of the raw data (dotted line). The peaks correspond to different photon multiplicities at each wavelength, with their mean values aligning with the data points in the calibration plot. The TES response is linear from 0 to \SI{5}{eV}, matching the operational range of the TES. Beyond \SI{5}{eV}, the response becomes nonlinear. A second-degree polynomial fit to the full energy spectrum is also shown.}
    \label{fig:Calibration_TESE}
\end{figure}

%**********************************************
To calibrate the TES, we have coupled it to three different attenuated monochromatic lasers of 1540, 850, and \SI{406}{nm} using an optical fiber placed about one millimeter away from the sensor.
The attenuation allows us to restrict the number of photons arriving at the TES and to avoid signal saturation, focusing on the region containing between one and eight photons.  
In Fig.~\ref{fig:Calibration_TESE}, the calibration curve is shown on the right, where we plot the measured signal amplitude from the TES against the known energy of the impinging photon(s). The left side of the figure displays the histograms of counts versus amplitude for each laser wavelength used.  
We have taken the absolute minimum value as the pulse amplitude from the unfiltered waveforms. Thus, even in the absence of energy deposition in the TES, the minimum value is nonzero, as fluctuations in the baseline signal around zero occur. This results in a nonzero intercept for the calibration curve. 
We have intentionally not applied any filtering during the calibration process in order to match the unfiltered oscilloscope conditions during data acquisition. 

%**********************************************
\section{Event classification}
%**********************************************
\begin{figure*}[t]
\centering
\includegraphics[width=0.9\textwidth, trim=1cm 1cm 1cm 0.98cm]
{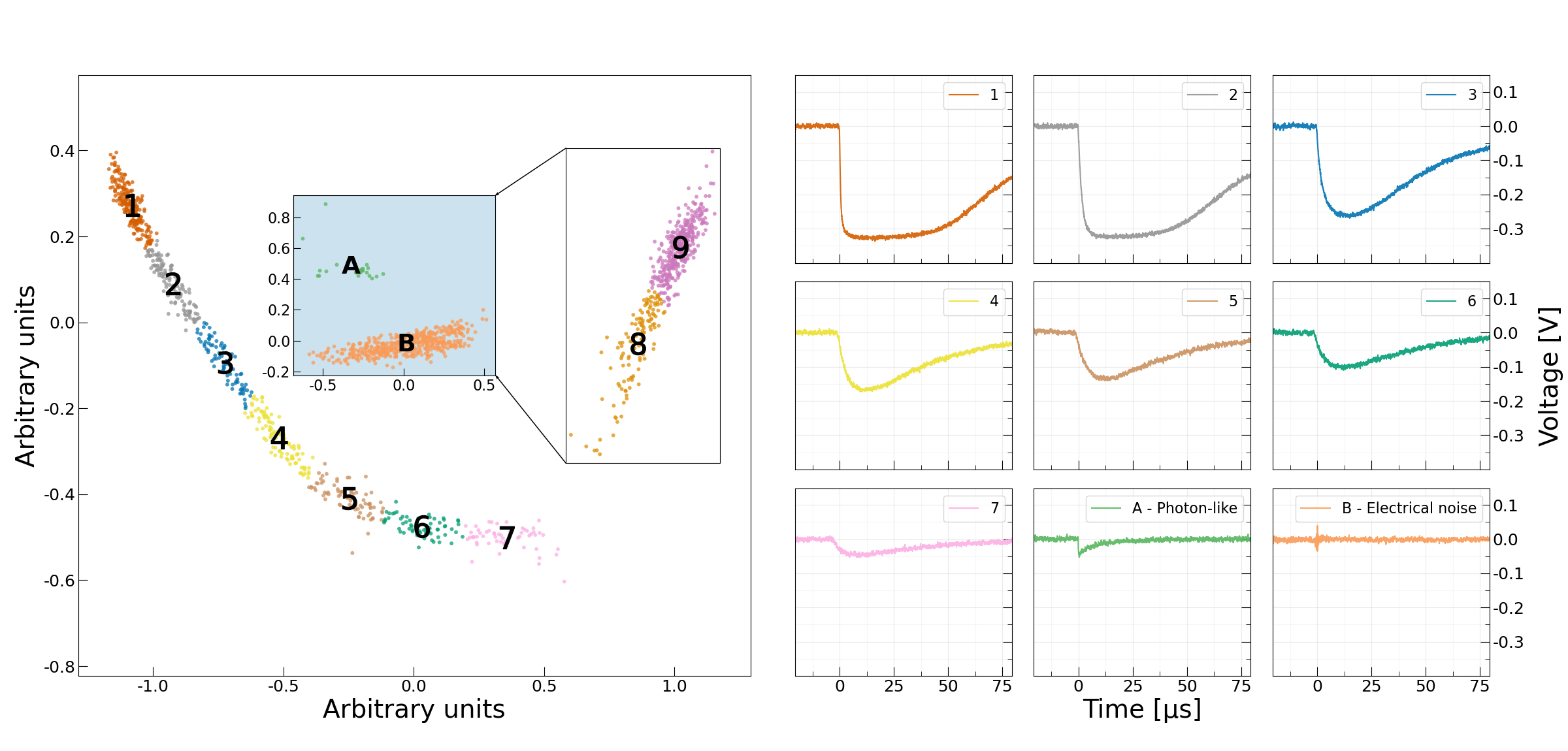}
\hfill
\caption{Left panel: K-means clustering of signals from background run after principal component analysis. The inset zooms in on clusters 8 and 9, which were reanalyzed to improve the separation of noise from photonlike events. Right panel: Sample signal waveforms from each of the refined clusters. For TES alone measurements like these, we recorded 2.5k sampling points at a rate of 25MHz for each oscilloscope trace, leading to a measurement window of \SI{100}{\micro s}.}
\label{fig:PCA}
\end{figure*} 
Signals detected by the TES without any input optics are referred to as ``intrinsic detector dark-counts'' by Ref.~\cite{dictionary}. However, in this paper we drop the adjective ``intrinsic,'' as we have identified external sources contributing to these counts such that not all of the dark-counts come from within the sensor (i.e., they are not intrinsic to it). The observed signals fall into three distinct categories: photonlike, high-energy, and electrical-noise events. While the three kinds of pulses are visually distinguishable, we have automated the categorization using the following pipeline developed in \textsc{python}. After applying a Butterworth filter, we calculate a set of variables for each waveform (e.g., full width at half maximum, maximum amplitude, pulse area, etc.). We employ principal component analysis (PCA) to find the two linear combinations of variables that maximize the variance of the points~\cite{wold1987pca}. $k$-means clustering is then applied to the resulting two-dimensional phase space to identify clusters of similar events~\cite{Arthur2007kmeans}. 
To ensure the accuracy of this automated sorting, we have conducted a manual review of the clusters.
This review process was facilitated by a graphical user interface (GUI) that allows for the selection of a cluster center, subsequently displaying all waveforms associated with that cluster for quick verification of the classification. In some cases, as demonstrated in Fig.~\ref{fig:PCA}, it was necessary to reapply the clustering algorithm on a subset of the data to fully separate the three groups. 

The photonlike events are the green points in group A in Fig.~\ref{fig:PCA}, which were obtained by rerunning the classification algorithm on clusters 8 and 9. These signals are characterized by a sharp rise time $\lesssim$\SI{1}{\micro s} and look identical to the ones we see when we shine light from the attenuated laser diode onto the TES.

The signals in clusters 1--7 in Fig.~\ref{fig:PCA} are characterized by a long rise and decay time, suggesting that these events are caused by large energy depositions in the TES substrate, likely coming from secondary cosmic ray particles or naturally occurring radioactive decays. We therefore refer to them as ``high-energy events.'' This classification is consistent with the observed long rise time, which can be attributed to the slow diffusion of energy from the substrate into the TES.

The electrical-noise events (orange cluster B in Fig.~\ref{fig:PCA} obtained by reanalyzing clusters 8 and 9) 
exhibit voltage spikes oscillating between positive and negative values and are characterized by a pulse area close to zero.
We attribute these events to electromagnetic interference with the readout apparatus. The fact that they are still detectable even when the TES is in a fully superconducting state, a condition in which it is not sensitive to energy depositions, indicates that the culprit is not the TES. 

%**********************************************
\section{Experimental Runs}
%**********************************************
The photonlike events are the expected signal for dielectric haloscopes searching for dark-matter dark photons. Therefore, it is essential to measure the photonlike DCR in the TES when the sensor is operated in the dark, with no optical link to ambient and in the absence of radioactive sources, as this forms the background against which the dielectric haloscope operates. 

\begin{figure}[t]
    \centering
    \includegraphics[width=0.8\linewidth]{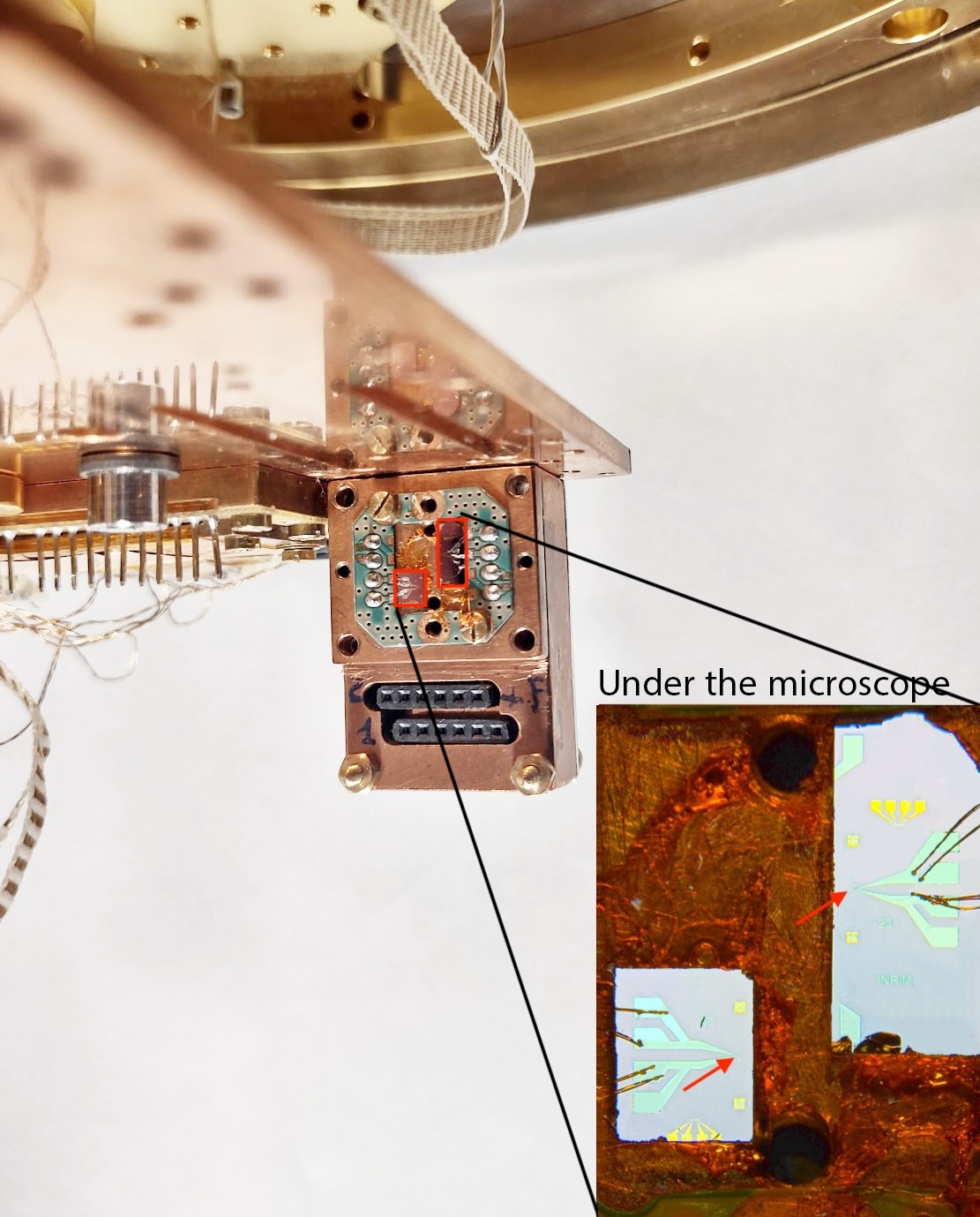}
    \caption[]{Attached vertically to the 30-mK-stage copper plate is the copper box---shown here without the lid---housing the two TESs. Although the TESs are too small to be seen, the substrates, highlighted by the red contours, are visible, with the top-right one (TES~E) almost double the size of the bottom-left one (TES~B). On the right is a microscope enlargement of the area within the green printed circuit board, with the TESs (still not visible) indicated by two red arrows.}
    \label{fig:photo}
\end{figure}

At the onset of our investigation, we expected to see a photonlike DCR on the order of $10^{-4}$\,Hz, in line with the findings by Ref.~\cite{dreyling2015characterization}, the only preceding study of a similar nature. There was also a question of whether cosmic rays or natural radioactive decays, interacting with either the TES or its substrate, could account for some of these photonlike events as suggested by the same authors. To explore this possibility, and to gain a deeper understanding of the nature of high-energy events, which had not been extensively studied in optical TESs before, we have carried out four specific tests, explained below. 
It is important to note that all the experimental runs were conducted in the absence of a light source and without an optical fiber connected to the TES, which was entirely encased in a copper box thermally coupled to the 30-mK-stage (see Fig.~\ref{fig:photo}).  
Although three fibers were present in the cryostat to allow for the initial calibration procedures, their ends were fixed to the outer side of the 3-K-stage within the ADR.

The first run was a background test (Run 1) intended to measure the photonlike DCR. Additionally, in the same run, we measured the high-energy DCR, which we hypothesized to be caused by cosmic rays or natural radioactive decays interacting with the TES substrate. 

In Runs 2 and 3, we used two radioactive sources, $^{232}$Th and $^{22}$Na, respectively.
These sources were placed against the exterior of the ADR and aligned perpendicularly to the TES substrate, which was mounted vertically inside the ADR (see Fig.~\ref{fig:photo}). 
The aim of these tests was to examine the response of the TES to gamma events using radioactive sources. We expected that environmental gamma would yield a similar response in the sensor.
In Run 2, a $^{232}$Th source with an activity measured at approximately \SI{10}{kBq} was used. For Run 3, we employed a $^{22}$Na source, which had an activity of around \SI{32}{kBq}. The emission spectrum of $^{22}$Na is characterized by two prominent photoelectric peaks. Notably, one of these peaks corresponds to the simultaneous emission of two back-to-back 511-keV photons, a phenomenon that occurs due to electron-positron annihilation following a $\beta^+$ decay. This source was sandwiched between the ADR and a plastic scintillator connected to a photomultiplier tube (PMT), which we call a ``saber.'' This setup allowed us to measure the double coincidences of the back-to-back gamma rays, capturing these occurrences with both the saber and the TES.

\begin{figure}[b]
    \centering
    \includegraphics[width=0.8\linewidth]{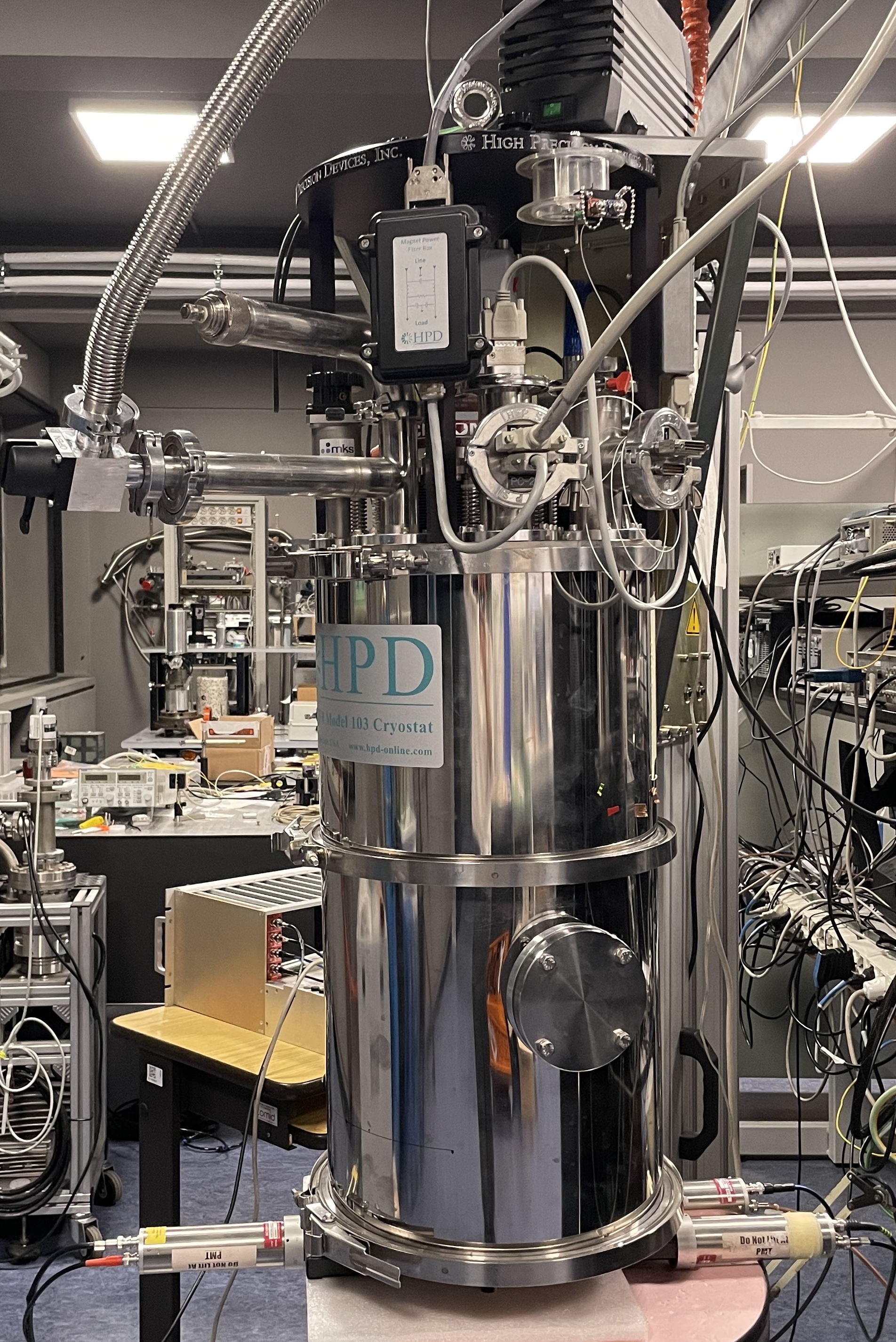}
    \caption[]{The experimental setup during the cosmic ray coincidence test. The three sabers, placed side by side, are directly below the ADR.}
    \label{fig:cryo_muon_coincidence}
\end{figure}

For the fourth test (Run 4), to demonstrate that cosmic rays can also induce high-energy events in the TES, we implemented a cosmic ray coincidence system, placing three sabers, side by side, directly below the ADR (see Fig.~\ref{fig:cryo_muon_coincidence}). A simultaneous triggering of the TES and all sabers indicated a cosmic ray shower or muon-bundle event. The three sabers were operated at the same gain within a 10\% uncertainty. The threshold was set at \SI{120}{mV} on all three, which corresponds to \SI{250}{keV} (for more details about the calibration of the sabers, see Appendix~\ref{appendix_saber_calibration}). 
The data acquisition was conducted in two modes: in Run~4A, which lasted about 19~hours, the oscilloscope was configured to trigger on the AND signal of the three sabers, recording the TES signal simultaneously. 
The coincidence in Run~4A was set using the smart trigger function of the oscilloscope, requiring a signal to exceed the threshold in all the saber channels across the entire \SI{10}{\micro s} window. In Run 4B, spanning around 14~hours, the trigger was set on the TES for signals greater than \SI{0.8}{eV}, while recording the outputs of all three sabers.  
Run 4A has provided the ratio of quadruple coincidences (events registered by both the TES and the three sabers) to triple coincidences (events detected solely by the sabers), while Run 4B allowed us to estimate the ratio of quadruple coincidences to the total number of high-energy events detected by the TES. 

%**********************************************
\section{GEANT4 Simulation}
%**********************************************
Preliminary calculations indicate that direct cosmic ray interactions with the TES alone cannot account for the observed number of high-energy events. To estimate the maximum possible direct cosmic ray contribution, we have considered a scenario in which the TES is oriented horizontally, maximizing its exposure. At sea level, the average muon flux is about \SI{120}{m^{-2}s^{-1}} (see, for example,~\cite{Arneodo_2019}). Even doubling this value to conservatively account for cosmic ray particles other than muons, a 20$\times$\SI{20}{\micro m^2} TES would receive only around $10^{-2}$ direct hits per day. In contrast, our larger 6.7$\times$\SI{2.7}{mm^2}, 500-µm-thick substrate would be impacted by around 400 cosmic ray particles per day. 
Since we observe about a thousand high-energy events per day in our measurements, far exceeding the estimated direct hits on the TES, we conclude that these events must (primarily) originate from energy depositions within substrate, and not direct interactions with the TES active area.
Indeed, as we shall see later, these high-energy events arise from both cosmic rays and environmental gamma rays interacting with the substrate.

%The substrate for TES~E is 6.7\,mm$\times$2.7\,mm in area and \SI{500}{\micro m} in thickness. 
We have developed a \textsc{geant4}-based simulation to validate our understanding of the high-energy events seen by the sensor. We have ran the simulation in three configurations called Sims 2, 3, and 4, which mirror the experimental setups of Runs 2, 3, and 4, respectively. All simulations have used the substrate as the sensitive detector rather than the TES itself.

In Sim 2, we modeled the $^{232}$Th source as in Run 2 and simulated one billion decay events. The simulation assumed the source to be in secular equilibrium. This approach is supported by empirical data obtained from analyzing the thorium source with a high-purity germanium (HPGe) detector at NYUAD, which provided activity-level and spectral information consistent with this assumption. The activity from these measurements, around \SI{10}{kBq}, was also used to convert the number of simulated decays to a time (i.e., approximately \SI{100000}{s} for \SI{1}{billion} decays).

In Sim 3, the $^{22}$Na source and the saber were positioned as in Run 3 and one billion decays were simulated. In the same way as for the thorium, we used the known activity of \SI{32}{kBq} to convert the simulated number of decays to a time. 

In Sim 4, we modeled the three sabers side by side, under the ADR, and generated ten billion secondary cosmic ray particles using the Cosmic-ray Shower Library (CRY package~\cite{Hagmann2007cry}). These cosmic ray particles, as well as those particles produced by their interactions with the setup, hit the substrate and the sabers. The saber hits were disregarded if their energy deposition was lower than \SI{250}{keV}, which corresponds to the trigger threshold used in Run~4A. 

In addition to cosmic rays, we also simulated environmental gamma-rays originating from the radioactivity of building materials and other naturally occurring radiation sources. These gamma rays were generated according to the actual gamma-ray spectrum measured in the laboratory, using a NaI crystal coupled to a photomultiplier tube. Our measurements indicated an average gamma flux of \numrange{1}{2}\,gamma cm$^{-2}$\,s$^{-1}$.

From Sim 4, we obtained a simulated high-energy DCR, which was then compared to the high-energy DCR observed in Run~1 (see the top green and orange points in  Fig.~\ref{fig:rate_BKG_Th_Na}). To accomplish this, we added together the rate of events detected in the TES substrate during the environmental gamma-ray simulation and the rate obtained during the cosmic ray simulation. To convert the simulated counts (the output from the simulation) to a simulated rate, we used the triple coincidence rate from Run 4A and the environmental gamma rate from the NaI detector as time-conversion factors. Additionally, Sim 4 enabled us to calculate the ratios of triple to quadruple coincidences and high-energy TES hits to quadruple coincidences. These ratios correspond to those measured in Runs 4A and 4B, respectively. 

\begin{figure}[tb]
    \centering
    \includegraphics[width=\linewidth]{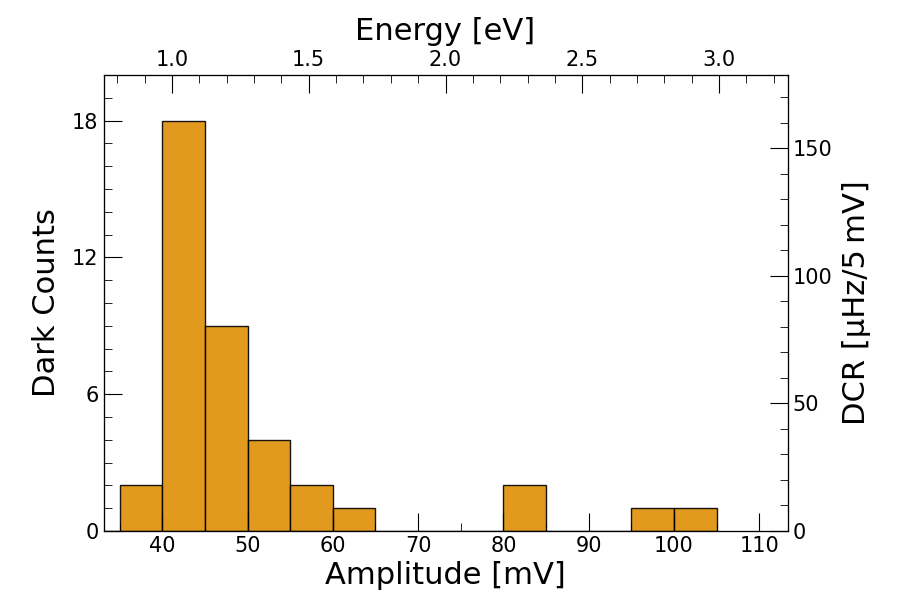}
    \caption[]{The histogram of the photonlike dark-counts versusthe amplitude (bottom \emph{x} axis) and the energy (top \emph{x} axis) during the background run: 40 photonlike events were recorded over 31 hours with a trigger threshold of \SI{0.8}{eV}. The right \emph{y} axis shows the rate in microhertz per bin. The 5-mV binning comes from the standard deviation of the baseline fluctuation.}
    \label{fig:DCR_TESE}
\end{figure}

\begin{figure*}[tb]
\centering
\includegraphics[width=0.9\textwidth, trim=1cm 0cm 1cm 0cm]
{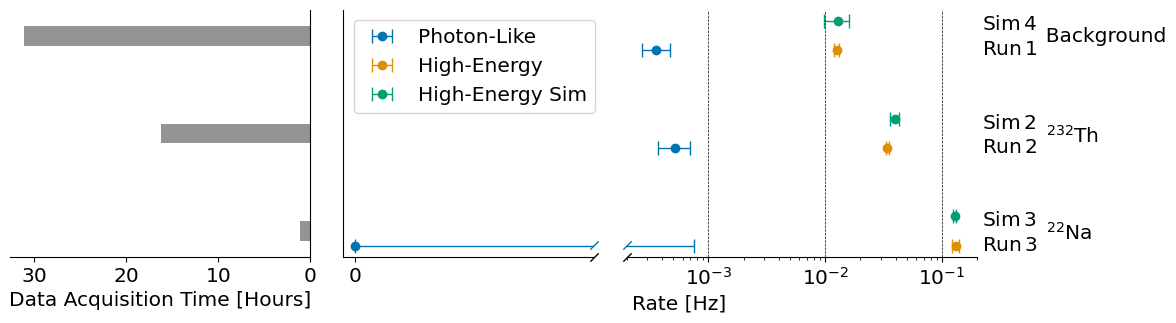}
\hfill
\caption{TES count rates under different experimental configurations, namely, when no radioactive source is present (background), with $^{232}$Th, and with $^{22}$Na. The measured photonlike rates are shown in blue, while the measured and simulated high-energy count rates are shown in orange and green, respectively. The photonlike rates could not be simulated due to their unknown origin. The background data come from Run 1, whereas the $^{232}$Th and $^{22}$Na data come from Runs 2 and 3. The simulated high-energy event rates come from Sims 4, 2, and 3, corresponding to cosmic rays and environmental gamma rays (which, together, are presumed to be the predominant cause of the high-energy background events), $^{232}$Th, and $^{22}$Na, respectively. As the simulations lack a built-in time, we have employed various time-scaling factors to convert simulation counts into rates. These factors are the triple coincidence rate from Run 4A, the environmental gamma rate from the NaI measurement, and the activities of thorium and sodium. The uncertainties associated with each measured data point represent the 90\% confidence interval from Poisson statistics. Given that we observed 0 counts in Run 3, we use Feldman and Cousins~\cite{Feldman1998Apr} confidence intervals for the error bar for that data point. For the simulated rates, the errors come from propagation of uncertainty in the time scaling factors. For the exact values of the points, see Table \ref{tab:sim_data} in appendix~\ref{appendix_table_sim_data}.}
\label{fig:rate_BKG_Th_Na}
\end{figure*}

%**********************************************
%\section{ICP-MS measurements}
%**********************************************
To investigate the contribution of radioactive decays to the number of high-energy events seen by the TES, we have conducted an inductively coupled plasma-mass spectrometry (ICP-MS) analysis at the Laboratori Nazionali del Gran Sasso (LNGS, L'Aquila, Italy). This was used to evaluate the intrinsic radioactivity of the materials around the TES, specifically the substrate and the copper box housing the TES. The analysis focused on uranium and thorium, which we suspected to be the main contributors of radioactive decays. In the substrate, the uranium and thorium concentrations were found to be less than \SI{3}{ppb} (parts per billion) and \SI{7}{ppb}, respectively. For the copper box, the concentrations of both isotopes were below \SI{1}{ppb}.
Upon incorporating the measured radioactivity of the copper box and substrate into our \textsc{geant4} simulation, it was determined that the TES would register approximately one hit every 500 days (\SI{21}{nHz}). This result indicates that the intrinsic radioactivity of the materials in close proximity to the TES contributes negligibly to its high-energy dark-count rate.

%**********************************************
\section{Results and discussion}
%**********************************************
The key finding from Run 1 is that the photonlike DCR is \num{3.6d-4}\,\unit{\hertz} within the energy range of 0.8--3.2\,eV. Notably, at $1.5\pm 0.2$\,eV---the energy range targeted by the MuDHI DM DP haloscope detector~\cite{manenti2022search}---the photonlike DCR is \num{6d-5}\,\unit{\hertz}. In Fig.~\ref{fig:DCR_TESE}, we show the photonlike DCR for the TES as a function of the photon energy and the signal amplitude.

\begin{figure}[b]
\centering
\includegraphics[width=\linewidth, trim={1.5cm 0 1.5cm 0}]{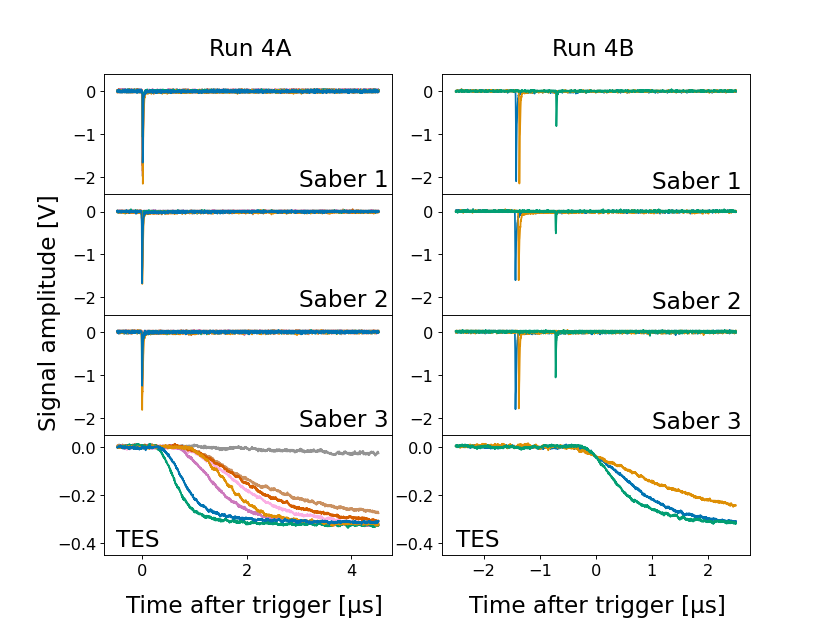}
\caption{Plots showing all quadruple coincidence events from Runs 4A (8 over 19 hours) and 4B (3 over 14 hours). For each simultaneous event detected across saber 1, 2, 3, and the TES, the same color is used in all four corresponding plots to represent that specific event. In Run 4A the trigger was on the AND signal of the three sabers. Due to the TES's longer, more variable rise time and delay relative to the sabers, the saber pulses appear to overlap. In Run 4B, the time window is extended to negative values to capture the saber pulses that occur prior to the TES trigger.}
\label{fig:Quad_Coincidences}
\end{figure}

Contrary to initial expectations, the photonlike DCR in the 0.8--3.2\,eV energy range remained constant (within uncertainties) across all runs (see the blue points in Fig.~\ref{fig:rate_BKG_Th_Na}), despite the introduction of radioactive sources in Runs 2 and 3. This consistency in the photonlike DCR, regardless of the presence of radioactive sources, suggests that gamma rays do not contribute to the observed photonlike event rate.
Despite including the Optical Physics package in our \textsc{geant4} simulation, no optical photons indicative of scintillation processes around the TESs were recorded.

One potential explanation for the observed photonlike DCR is the presence of stray optical photons within the cryostat, possibly from external ambient light. However, the likelihood of these photons entering through the optical fiber seems minimal, as any ambient light entering the fiber would be absorbed by the 3-K stage to which it is fixed. Supporting this, previous research indicates that the presence or absence of a fiber fed into the cryostat, but not directly coupled to the TES, has no significant effect on the photonlike DCR (see Fig.~4.13 in Ref.~\cite{dreyling2014superconducting}).
This suggests that the source of these photonlike signals remains unknown, hence the absence of simulation points in Fig.~\ref{fig:rate_BKG_Th_Na} for the photonlike DCR, and requires further investigation.
 
In Run 1, we selected high-energy signals to calculate the high-energy dark-count rate. This rate is caused by cosmic ray and environmental gamma-ray interactions with the TES substrate, as confirmed by our simulations.
Notably, for TES~B, the substrte area of which is half the size of that of TES~E, we observed that the high-energy DCR was half that of TES~E, substantiating the relationship between the size of the substrate and the high-energy DCR (for more details, see Appendix~\ref{appendix_sensitive volume}). 

Upon introducing radioactive sources, a noticeable increase in the rate of high-energy signals in the TES was observed on the oscilloscope. This was particularly visible with the $^{22}$Na source. Our simulation shows that we can accurately model the high-energy events in the presence of a radioactive source, with compatible results for both the $^{232}$Th and $^{22}$Na run. The simulation indicates that the substrate absorbs the energy and then transmits it to the TES, where it is detected. Notably, altering the substrate area linearly changes the DCR, further confirming this interaction mechanism.

Run 4 provided confirmation that cosmic ray showers are capable of producing high-energy events in the TES, as evidenced by the quadruple coincidences observed between the sensor and the three sabers below the ADR (see Fig.~\ref{fig:Quad_Coincidences}).
The simulated ratio of triple coincidences to quadruple coincidences, which represents the fraction of cosmic ray events detected by the saber system that also deposited energy in the TES, overestimates the one observed in Run 4A by a factor of 5. Similarly, the simulated ratio of high-energy TES hits to quadruple coincidences overestimates Run 4B by a factor of 1.7. These discrepancies are likely due to systematics in the cosmic ray simulation. Table~\ref{tab:Run4} shows the raw counts for triple and quadruple coincidences in simulation and real life.

\begin{table}[t]
    \caption{The table compares the ratio of triple (t) to quadruple (q) coincidence events between simulation (Sim) and experimental data (Exp) for Run 4A, and the ratio of high-energy TES hits (h) to quadruple between simulation and experimental data for Run 4B. In Run 4A, the simulation predicts a significantly higher triple to quadruple ratio, approximately 5 times larger than observed in the experimental data. In Run 4B, the simulation overestimates the ratio of high-energy TES hits to quadruple coincidences, albeit to a smaller extent, predicting a value 1.7 times higher than measured experimentally.}
    \label{tab:Run4}    
    \setlength{\tabcolsep}{8pt}
    \def\arraystretch{1.3}%
    \begin{tabular}{l | l l l }
                 &   Sim         &   Exp           & Ratio Sim/Exp   \\ \hline
        Run~4A t/q  &   \num{107524}/46   &   3818/8     &     5        \\ 
        Run~4B h/q  &   \num{24906}/46    &   963/3      &   1.7          \\ 
    \end{tabular}
\end{table}

\section{Conclusions}
We have characterized the dark-count rate of a Ti/Au superconducting transition-edge sensor. By employing singular-value decomposition, principal component analysis, and \emph{k}-means clustering, we have effectively filtered out noise and high-energy events to isolate photonlike signals. These photonlike signals are of particular interest since they resemble the expected dark photon signal in a dielectric haloscope. We have achieved a photonlike DCR of \num{3.6d-4}\,Hz in the \numrange{0.8}{3.2}\,eV range. Removing optical connections between the TES and ambient is essential to achieve very low DCRs. The source of the observed residual photonlike events still remains unexplained, necessitating further investigation. 

Regarding high-energy events, our runs using $^{232}$Th and $^{22}$Na sources, along with coincidence measurements between the TES and an external cosmic ray detector, coupled with a detailed \textsc{geant4} simulation, have allowed us to ascribe these events to high-energy particle impacts in the TES substrate. Previous studies hypothesized that these high-energy events are due to radiation and cosmic ray interactions with the substrate, but have lacked experimental and simulation verification. Additionally, natural radiation from materials in close proximity to the TES appears to have a negligible impact on the high-energy DCR of the TES. Instead, the primary contributors to this rate are cosmic ray particles and environmental gamma rays hitting the substrate. All code used for the simulations is available on our GitHub repository~\cite{github}.

\section*{Acknowledgments}
Part of this work has been carried out at QR Lab-Micro\&Nanolaboratories, Istituto Nazionale di Ricerca Metrologica (INRiM).
We thank Stefano Nisi and Francesco Ferella at Laboratori Nazionali del Gran Sasso (LNGS) for the ICP-MS measurements. 
This research was carried out on the High Performance Computing resources at New York University Abu Dhabi.
We are grateful to the NYUAD Kawader Research Assistantship Program. This work was also made possible by the contribution of Grant (62313) from the John Templeton Foundation. Fermilab is operated by Fermi Research Alliance, LLC under Contract
No. DE-AC02-07CH11359 with the United States Department of Energy.
%\bibliography{bibliography}% Produces the bibliography via BibTeX.
%

\section*{Appendix}
\appendix

%*******************************
\section{TES fabrication}
\label{appendix_TES_fabrication}
%*******************************
Both TES~B and TES~E were fabricated at INRiM. They are composed of a 15-nm film of Ti under a 30-nm film of Au deposited on a substrate~\cite{monticone2021ti}. The substrate was made through low-pressure chemical vapor deposition (LPCVC) and is composed of a Si substrate (525$\pm$\SI{25}{\micro m} thick), sandwiched between two bilayers composed of SiN$_{x}$ silicon nitride (\SI{500}{nm} thick) and SiO$_2$ thermal oxide (\SI{150}{nm} thick), with the silicon nitride as the outside layer.

The depositions were performed in a custom ultrahigh-vacuum (UHV) deposition chamber at a pressure less than \num{5d-6}\,\unit{\Pa}, which is essential to prevent Ti oxidation and achieve a smooth interface between titanium and gold. 
The outlines of the TESs were defined by the lift-off technique using a Heidelberg \textmu PG101 laser writer.

Before the deposition, the substrate underwent sputter etching for 15\,s at \SI{400}{V}. Then, the sample was moved to the UHV chamber for the Ti and Au deposition. This consisted of Ti deposition by $e$ beam at a rate of \SI{0.25}{nm/s}, followed, after about \SI{30}{s}, by the deposition of Au by effusion cell at a rate of \SI{0.07}{nm/s}. The deposited thicknesses were continually monitored by a quartz-crystal sensor.

The wiring was made by lift-off of \SI{50}{nm} of sputtered Nb. Sputter etching was used before the wiring process to reduce the contact resistance between the Nb and the TESs.

%*******************************
\section{Classification algorithm}
\label{appendix_classification}
%*******************************
To classify the event type of each waveform, we first apply the SciPy Butterworth filter with low-pass mode, order 2, frequency 1/30, which, when applied to the TES triggered data sampled at \SI{25}{MHz}, corresponds to \SI{417}{kHz}, and we then calculate a set of variables including the maximum amplitude, the area, the full-width at half maximum, the standard deviation, the average positive and negative gradient, and the cross-correlation with a reference photonlike event. This gives us a point in an \emph{n}-dimensional space for each TES pulse recorded by the oscilloscope.

We then perform a PCA on the matrix containing the aforementioned n-dimensional points~\cite{wold1987pca}. Using this technique, we can identify the two axes in this n-dimensional space that maximize the variance among these points. Plotting pulses on these two axes typically results in distinct clusters. This representation enables the visual differentiation of photonlike, high-energy, and electrical-noise events. 

Furthermore, applying \emph{k}-means clustering allows us to easily collate pulses that fall close together in the n-dimensional space~\cite{Arthur2007kmeans}. The combined use of these tools, and, if necessary, reanalysis of selected pulse subgroups, enables efficient pulse-shape discrimination on datasets comprised of thousands to tens of thousands of pulses in approximately 10\,min. 

The software was developed in Python using the SciPy module~\cite{2020SciPy} and it is available at Ref.~\cite{github2}. 

%*******************************
\section{GEANT4}
\label{appendix_GEANT4}
%*******************************
We used the \textsc{geant4} v.11.1.2 software, an object-oriented framework written in C++ that employs Monte Carlo methods~\cite{agostinelli2003geant4}. In our simulations, we activated the EmLiverMore low-energy physics module, with the particle-induced x-ray Emission (PIXE) parameter enabled. 

During each simulation event, data were collected on particle characteristics, including the position, the initial energy, the primary vertex, the sensitive volume or volumes that it hits, and the energy deposited. To simulate cosmic ray particles, which include neutrons, protons, gamma rays, electrons, pions, and muons, we used the Cosmic Ray Shower Library (CRY)~\cite{Hagmann2007cry}. To simulate the environmental radioactivity we used the gamma spectrum measured in the laboratory, by means of a NaI detector. The gamma rays were generated isotropically on a cylindrical surface surrounding the ADR. 

\begin{figure}[htbp]
    \centering
    \includegraphics[width=\linewidth]{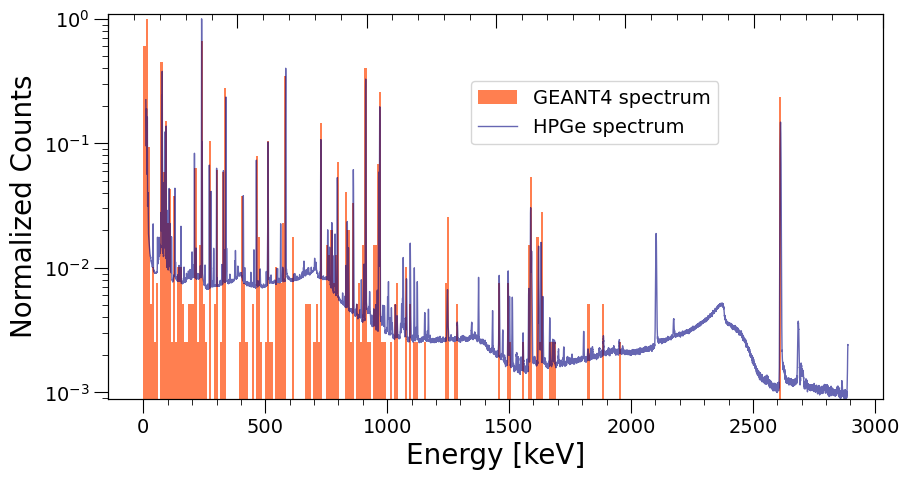}
    \caption[]{Comparison of the $^{232}$Th spectra: \textsc{geant4} simulation and measurement from the HPGe detector of the thorium source. \textsc{geant4} counts are scaled down for easy comparison with the measured counts.}
    \label{fig:SimulationSpectrumArgument}
\end{figure}

For simulations involving thorium and sodium, we employed the Radioactive Decay module along with the General Particle Source. To optimize the computing time, we limited our analysis to secondary products with a Track ID of 50 or less. We validated this approach by comparing the $^{232}$Th spectrum from our simulation with measurements from the High Purity Germanium detector at NYUAD (Fig.~\ref{fig:SimulationSpectrumArgument}). 

The geometry of the detector was defined using Geometry Description Markup Language (GDML) files, which, when combined with macro files, allowed for the dynamic adjustment of various simulation parameters. For computationally intensive simulations, we used the message passing interface (MPI) on a high-performance computer, which allowed us to use approximately 8000 CPU hours per simulation.

%*******************************
\section{TES~B}
\label{appendix_TESB}
\begin{figure}[tb]
    \centering
    \includegraphics[width=\linewidth]{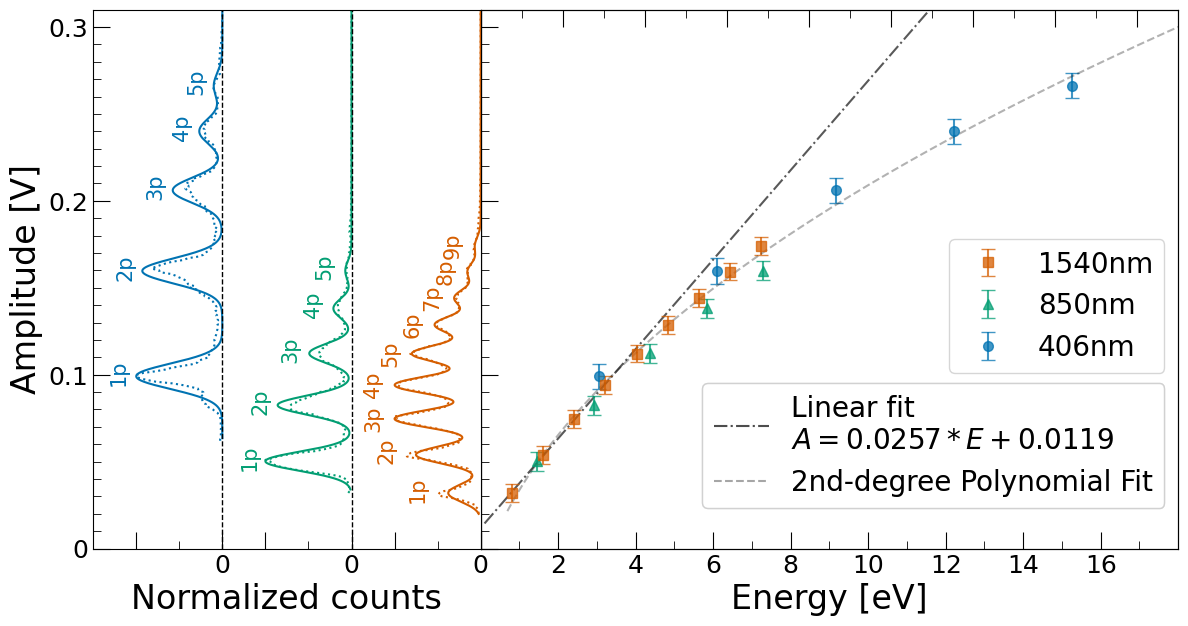}
    \caption[]{The energy calibration of TES~B.}
    \label{fig:Calibration_TESB}
\end{figure}

%*******************************
In Fig.~\ref{fig:Calibration_TESB}, we show the energy-calibration curve for TES~B.

A background run of 14 hours and 45 minutes resulted in seven photonlike counts and 370 high-energy counts. 
The high-energy dark-count rate is half that of TES~E, the substrate of which is twice the size of that of TES~B. This is consistent with the hypothesis that high-energy events are caused by cosmic ray particle interactions within the substrate.
However, the photonlike DCR is effectively the same as that of TES~E, implying that the photonlike events do not originate in the substrate and, as such, are unaffected by the difference in substrate size.

The fact that the results from both TESs are consistent indicates that the fabrication techniques that we employ are reliable and lead to reproducible sensors. 

%*******************************
\section{Sensitive volume}
\label{appendix_sensitive volume}
\begin{figure}[htbp]
    \centering
    \includegraphics[width=0.8\linewidth]{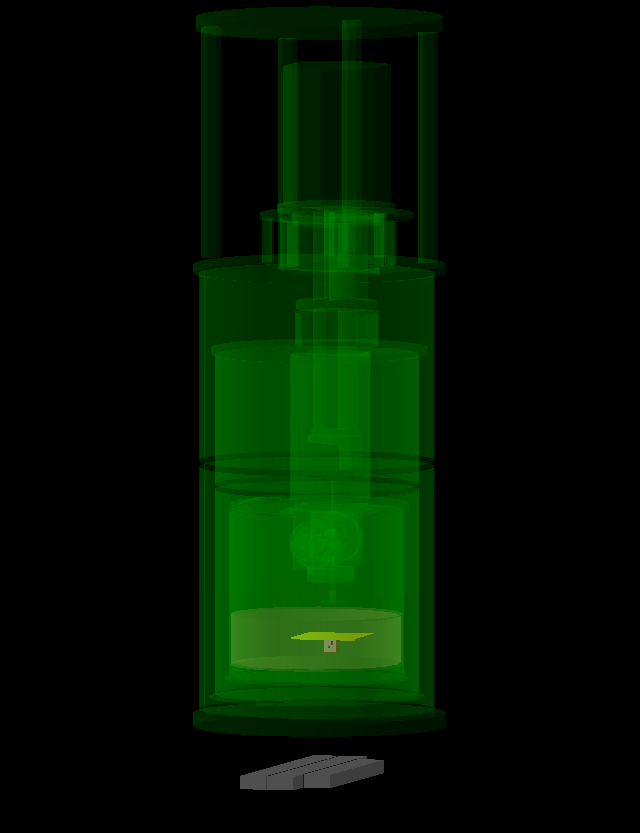}
    \caption[]{The \textsc{geant4 GUI} output of Sim 4. It includes the ADR, the 30-mK plate, the magnetic shield, the copper box housing the TES substrates, and, in gray, the three sabers below the refrigerator.}
    \label{fig:GEANT4_Snapshot}
\end{figure}

%*******************************
To assess whether the TESs were sensitive just to energy depositions in the substrate or also to energy depositions in the base of the copper box hosting them, a thorium source experiment involving both TESs was conducted. A \textsc{geant4} simulation was also performed (see Fig.~\ref{fig:GEANT4_Snapshot}).

\begin{table}[b]
    \caption{The data from a thorium run with the two TESs operating simultaneously over 17.5 hours and a comparison with simulation.}
    \label{tab:ThoriumSummerRun}    
    \setlength{\tabcolsep}{10pt}
    \def\arraystretch{1.3}%
    \begin{tabular}{l | l l l l l}
                     & TES~E       & TES~B       & Both   \\ \hline
        Experiment   &   687       &   464       &   7          \\ 
        Simulation   &   563       &   472       &   5          \\ 
    \end{tabular}
\end{table}

Our results are summarized in Table~\ref{tab:ThoriumSummerRun}.
Both the experimental data and the simulation show comparable ratios between the TES~E and TES~B counts, as well as similar numbers of double coincidences. These results reinforce our belief that the substrate, which is the sensitive volume in the simulation, is the only volume where energy depositions lead to detections in the TESs. 

Given that the base of the copper box is not a sensitive volume in the simulation, if energy deposited in the copper could be detected by the TESs in real life, we would expect the simulation to underestimate the TES hits. Furthermore, since the base of the copper box connects both TESs, if the copper were able to diffuse heat to the sensors, we would observe significantly more double coincidences in real life than in simulation. Since we did not observe either, we conclude that energy deposited in the copper does not contribute to TES hits. 

To investigate further the nature of the double coincidences, we have calculated the rate of random coincidences using the formula $N_{\rm EB}=2N_{\rm E}N_{\rm B}\cdot\frac{T}{t}$, where $N_{\rm EB}$ is the number of double coincidences, $N_{\rm E}$ and $N_{\rm B}$ are the numbers of hits on TES~E and TES~B, respectively, T is the gate time, and t is the total measurement time. For our specific setup, which included a gate time of \SI{100}{\micro s} and a total time of 17.5~hours, the expected number of random double coincidences is 0.001. This number is markedly lower than the seven occurrences that we observed over the 17.5 hours. This confirms that the double coincidences are correlated and likely caused by the same initial $^{232}$Th decay.
Through simulation, it was understood that the main source of such coincidences is secondary electrons, originating in the copper box from $^{232}$Th decay events, impacting the substrates simultaneously. 

To further validate that the energy depositions causing the high-energy events occur only within the substrate, we have conducted a series of simulations varying both the size of the substrate and its distance from the radioactive source. These simulations conform best to reality when the sensitive volume matches the real substrate size and position and they reveal a square-law relationship between the count rate and the distance from the radioactive source.

\begin{figure}[tb]
\centering
\includegraphics[width=\linewidth]{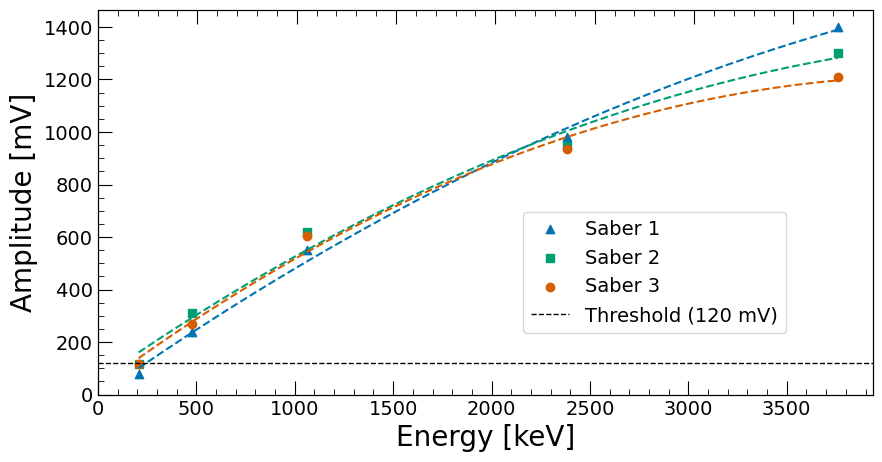}
\caption[]{The energy calibration for the three sabers at similar PMT gains. The trigger on the oscilloscope during data acquisition was set to \SI{120}{mV}. The lines correspond to the second-order polynomial fits of the data points for each saber.}
\label{fig:Saber_calibration}
\end{figure}

\section{Saber calibration}
\label{appendix_saber_calibration}

The calibration of the sabers was carried out at NYUAD using different radioactive sources. 
For each source, we identified the characteristic Compton edge and measured the corresponding peak height in the detected amplitude spectrum (in millivolts). In Fig.~\ref{fig:Saber_calibration}, we show the calibration curve for all three sabers with the PMTs operating at similar gains, while in Table~\ref{tab:pmt_calibration}, we list the raw data.

\begin{table*}[t]
\caption{Calibration data from the sabers using various radioactive sources.  Due to the lower probability of the photoelectric effect occurring compared to Compton scattering in plastic scintillators, we use the Compton edge for saber calibration.}

\setlength{\tabcolsep}{10pt}
\def\arraystretch{1.3}%
\begin{tabular}{c|l|l|l|l}

         & Saber 1 [mV] & Saber 2 [mV] & Saber 3 [mV] %& Peak [keV]
         & Compton edge [keV]  \\ \hline
$^{133}$Ba   & 78      & 115     & 116     %& 356        
& 207           \\ 
$^{137}$Cs   & 240     & 310     & 270     %& 662        
& 478           \\ 
$^{22}$Na    & 550     & 620     & 604     %& 1275       
& 1062          \\ 
$^{232}$Th   & 980     & 945     & 935     %& 2614       
& 2381          \\ 
Muons    & 1400    & 1300    & 1210    %& 4000       
& 3760           
\end{tabular}
\label{tab:pmt_calibration}
\end{table*}

%*******************************
\section{Simulation and experimental data}
\label{appendix_table_sim_data}
%*******************************
In Table \ref{tab:sim_data}, we report the values obtained in the simulation and in real life for the photonlike and the high-energy DCR during the background, thorium, and sodium runs.
\begin{table*}[t]
\caption{The values of the plot shown in Fig.~\ref{fig:rate_BKG_Th_Na} along with their uncertainties.}

\setlength{\tabcolsep}{10pt}
\def\arraystretch{1.3}%
\begin{tabular}{c|l|l|l}

            & Photonlike [mHz]  & High-energy [mHz]  & High-energy Sim [mHz] \\ \hline
Background  & 0.36 $^{+0.11}_{-0.09}$     
& 12.10 $^{+0.56}_{-0.54}$     
& 12.83 $^{+0.39}_{-0.33}$                \\ 
$^{232}$Th  & 0.52 $^{+0.18}_{-0.14}$    
& 34.00 $^{+1.28}_{-1.25}$    
& 39.20 $^{+3.44}_{-3.44}$               \\ 
$^{22}$Na   & 0.00 $^{+0.75}_{-0.00}$             
& 131.00 $^{+9.79}_{-9.25}$     
& 129.00 $^{+4.02}_{-4.02}$                
\end{tabular}
\label{tab:sim_data}
\end{table*}

\clearpage
\bibliographystyle{apsrev4-2}
\bibliography{bibliography}

\begin{thebibliography}{25}%
\makeatletter
\providecommand \@ifxundefined [1]{%
 \@ifx{#1\undefined}
}%
\providecommand \@ifnum [1]{%
 \ifnum #1\expandafter \@firstoftwo
 \else \expandafter \@secondoftwo
 \fi
}%
\providecommand \@ifx [1]{%
 \ifx #1\expandafter \@firstoftwo
 \else \expandafter \@secondoftwo
 \fi
}%
\providecommand \natexlab [1]{#1}%
\providecommand \enquote  [1]{``#1''}%
\providecommand \bibnamefont  [1]{#1}%
\providecommand \bibfnamefont [1]{#1}%
\providecommand \citenamefont [1]{#1}%
\providecommand \href@noop [0]{\@secondoftwo}%
\providecommand \href [0]{\begingroup \@sanitize@url \@href}%
\providecommand \@href[1]{\@@startlink{#1}\@@href}%
\providecommand \@@href[1]{\endgroup#1\@@endlink}%
\providecommand \@sanitize@url [0]{\catcode `\\12\catcode `\$12\catcode `\&12\catcode `\#12\catcode `\^12\catcode `\_12\catcode `\%12\relax}%
\providecommand \@@startlink[1]{}%
\providecommand \@@endlink[0]{}%
\providecommand \url  [0]{\begingroup\@sanitize@url \@url }%
\providecommand \@url [1]{\endgroup\@href {#1}{\urlprefix }}%
\providecommand \urlprefix  [0]{URL }%
\providecommand \Eprint [0]{\href }%
\providecommand \doibase [0]{https://doi.org/}%
\providecommand \selectlanguage [0]{\@gobble}%
\providecommand \bibinfo  [0]{\@secondoftwo}%
\providecommand \bibfield  [0]{\@secondoftwo}%
\providecommand \translation [1]{[#1]}%
\providecommand \BibitemOpen [0]{}%
\providecommand \bibitemStop [0]{}%
\providecommand \bibitemNoStop [0]{.\EOS\space}%
\providecommand \EOS [0]{\spacefactor3000\relax}%
\providecommand \BibitemShut  [1]{\csname bibitem#1\endcsname}%
\let\auto@bib@innerbib\@empty
%</preamble>
\bibitem [{\citenamefont {Irwin}(1995)}]{irwin1995application}%
  \BibitemOpen
  \bibfield  {author} {\bibinfo {author} {\bibfnamefont {K.}~\bibnamefont {Irwin}},\ }\href {https://doi.org/10.1063/1.113674} {\bibfield  {journal} {\bibinfo  {journal} {Applied Physics Letters}\ }\textbf {\bibinfo {volume} {66}},\ \bibinfo {pages} {1998} (\bibinfo {year} {1995})}\BibitemShut {NoStop}%
\bibitem [{\citenamefont {Cocco}\ \emph {et~al.}(2007)\citenamefont {Cocco}, \citenamefont {Mangano},\ and\ \citenamefont {Messina}}]{Cocco2007Jun}%
  \BibitemOpen
  \bibfield  {author} {\bibinfo {author} {\bibfnamefont {A.~G.}\ \bibnamefont {Cocco}}, \bibinfo {author} {\bibfnamefont {G.}~\bibnamefont {Mangano}},\ and\ \bibinfo {author} {\bibfnamefont {M.}~\bibnamefont {Messina}},\ }\href {https://doi.org/10.1088/1475-7516/2007/06/015} {\bibfield  {journal} {\bibinfo  {journal} {Journal of Cosmology and Astroparticle Physics}\ }\textbf {\bibinfo {volume} {2007}}\bibinfo  {number} { (06)},\ \bibinfo {pages} {015}}\BibitemShut {NoStop}%
\bibitem [{\citenamefont {{PTOLEMY Collaboration}}\ \emph {et~al.}(2022)\citenamefont {{PTOLEMY Collaboration}}, \citenamefont {Apponi}, \citenamefont {Betti}, \citenamefont {Borghesi}, \citenamefont {Boyarsky}, \citenamefont {Canci}, \citenamefont {Cavoto}, \citenamefont {Chang}, \citenamefont {Cheianov}, \citenamefont {Cheipesh}, \citenamefont {Chung}, \citenamefont {Cocco}, \citenamefont {Colijn}, \citenamefont {D{'}Ambrosio}, \citenamefont {de~Groot}, \citenamefont {Esposito}, \citenamefont {Faverzani}, \citenamefont {Ferella}, \citenamefont {Ferri}, \citenamefont {Ficcadenti}, \citenamefont {Frederico}, \citenamefont {Gariazzo}, \citenamefont {Gatti}, \citenamefont {Gentile}, \citenamefont {Giachero}, \citenamefont {Hochberg}, \citenamefont {Kahn}, \citenamefont {Lisanti}, \citenamefont {Mangano}, \citenamefont {Marcucci}, \citenamefont {Mariani}, \citenamefont {Marques}, \citenamefont {Menichetti}, \citenamefont {Messina}, \citenamefont {Mikulenko}, \citenamefont {Monticone}, \citenamefont {Nucciotti},
  \citenamefont {Orlandi}, \citenamefont {Pandolfi}, \citenamefont {Parlati}, \citenamefont {Pepe}, \citenamefont {P{\ifmmode\acute{e}\else\'{e}\fi}rez de~los Heros}, \citenamefont {Pisanti}, \citenamefont {Polini}, \citenamefont {Polosa}, \citenamefont {Puiu}, \citenamefont {Rago}, \citenamefont {Raitses}, \citenamefont {Rajteri}, \citenamefont {Rossi}, \citenamefont {Rozwadowska}, \citenamefont {Rucandio}, \citenamefont {Ruocco}, \citenamefont {Strid}, \citenamefont {Tan}, \citenamefont {Teles}, \citenamefont {Tozzini}, \citenamefont {Tully}, \citenamefont {Viviani}, \citenamefont {Zeitler},\ and\ \citenamefont {Zhao}}]{PTOLEMYCollaboration2022Sep}%
  \BibitemOpen
\bibfield  {number} {  }\bibfield  {author} {\bibinfo {author} {\bibnamefont {{PTOLEMY Collaboration}}}, \bibinfo {author} {\bibfnamefont {A.}~\bibnamefont {Apponi}}, \bibinfo {author} {\bibfnamefont {M.~G.}\ \bibnamefont {Betti}}, \bibinfo {author} {\bibfnamefont {M.}~\bibnamefont {Borghesi}}, \bibinfo {author} {\bibfnamefont {A.}~\bibnamefont {Boyarsky}}, \bibinfo {author} {\bibfnamefont {N.}~\bibnamefont {Canci}}, \bibinfo {author} {\bibfnamefont {G.}~\bibnamefont {Cavoto}}, \bibinfo {author} {\bibfnamefont {C.}~\bibnamefont {Chang}}, \bibinfo {author} {\bibfnamefont {V.}~\bibnamefont {Cheianov}}, \bibinfo {author} {\bibfnamefont {Y.}~\bibnamefont {Cheipesh}}, \bibinfo {author} {\bibfnamefont {W.}~\bibnamefont {Chung}}, \bibinfo {author} {\bibfnamefont {A.~G.}\ \bibnamefont {Cocco}}, \bibinfo {author} {\bibfnamefont {A.~P.}\ \bibnamefont {Colijn}}, \bibinfo {author} {\bibfnamefont {N.}~\bibnamefont {D{'}Ambrosio}}, \bibinfo {author} {\bibfnamefont {N.}~\bibnamefont {de~Groot}}, \bibinfo {author}
  {\bibfnamefont {A.}~\bibnamefont {Esposito}}, \bibinfo {author} {\bibfnamefont {M.}~\bibnamefont {Faverzani}}, \bibinfo {author} {\bibfnamefont {A.}~\bibnamefont {Ferella}}, \bibinfo {author} {\bibfnamefont {E.}~\bibnamefont {Ferri}}, \bibinfo {author} {\bibfnamefont {L.}~\bibnamefont {Ficcadenti}}, \bibinfo {author} {\bibfnamefont {T.}~\bibnamefont {Frederico}}, \bibinfo {author} {\bibfnamefont {S.}~\bibnamefont {Gariazzo}}, \bibinfo {author} {\bibfnamefont {F.}~\bibnamefont {Gatti}}, \bibinfo {author} {\bibfnamefont {C.}~\bibnamefont {Gentile}}, \bibinfo {author} {\bibfnamefont {A.}~\bibnamefont {Giachero}}, \bibinfo {author} {\bibfnamefont {Y.}~\bibnamefont {Hochberg}}, \bibinfo {author} {\bibfnamefont {Y.}~\bibnamefont {Kahn}}, \bibinfo {author} {\bibfnamefont {M.}~\bibnamefont {Lisanti}}, \bibinfo {author} {\bibfnamefont {G.}~\bibnamefont {Mangano}}, \bibinfo {author} {\bibfnamefont {L.~E.}\ \bibnamefont {Marcucci}}, \bibinfo {author} {\bibfnamefont {C.}~\bibnamefont {Mariani}}, \bibinfo {author}
  {\bibfnamefont {M.}~\bibnamefont {Marques}}, \bibinfo {author} {\bibfnamefont {G.}~\bibnamefont {Menichetti}}, \bibinfo {author} {\bibfnamefont {M.}~\bibnamefont {Messina}}, \bibinfo {author} {\bibfnamefont {O.}~\bibnamefont {Mikulenko}}, \bibinfo {author} {\bibfnamefont {E.}~\bibnamefont {Monticone}}, \bibinfo {author} {\bibfnamefont {A.}~\bibnamefont {Nucciotti}}, \bibinfo {author} {\bibfnamefont {D.}~\bibnamefont {Orlandi}}, \bibinfo {author} {\bibfnamefont {F.}~\bibnamefont {Pandolfi}}, \bibinfo {author} {\bibfnamefont {S.}~\bibnamefont {Parlati}}, \bibinfo {author} {\bibfnamefont {C.}~\bibnamefont {Pepe}}, \bibinfo {author} {\bibfnamefont {C.}~\bibnamefont {P{\ifmmode\acute{e}\else\'{e}\fi}rez de~los Heros}}, \bibinfo {author} {\bibfnamefont {O.}~\bibnamefont {Pisanti}}, \bibinfo {author} {\bibfnamefont {M.}~\bibnamefont {Polini}}, \bibinfo {author} {\bibfnamefont {A.~D.}\ \bibnamefont {Polosa}}, \bibinfo {author} {\bibfnamefont {A.}~\bibnamefont {Puiu}}, \bibinfo {author} {\bibfnamefont
  {I.}~\bibnamefont {Rago}}, \bibinfo {author} {\bibfnamefont {Y.}~\bibnamefont {Raitses}}, \bibinfo {author} {\bibfnamefont {M.}~\bibnamefont {Rajteri}}, \bibinfo {author} {\bibfnamefont {N.}~\bibnamefont {Rossi}}, \bibinfo {author} {\bibfnamefont {K.}~\bibnamefont {Rozwadowska}}, \bibinfo {author} {\bibfnamefont {I.}~\bibnamefont {Rucandio}}, \bibinfo {author} {\bibfnamefont {A.}~\bibnamefont {Ruocco}}, \bibinfo {author} {\bibfnamefont {C.~F.}\ \bibnamefont {Strid}}, \bibinfo {author} {\bibfnamefont {A.}~\bibnamefont {Tan}}, \bibinfo {author} {\bibfnamefont {L.~K.}\ \bibnamefont {Teles}}, \bibinfo {author} {\bibfnamefont {V.}~\bibnamefont {Tozzini}}, \bibinfo {author} {\bibfnamefont {C.~G.}\ \bibnamefont {Tully}}, \bibinfo {author} {\bibfnamefont {M.}~\bibnamefont {Viviani}}, \bibinfo {author} {\bibfnamefont {U.}~\bibnamefont {Zeitler}},\ and\ \bibinfo {author} {\bibfnamefont {F.}~\bibnamefont {Zhao}},\ }\href {https://doi.org/10.1103/PhysRevD.106.053002} {\bibfield  {journal} {\bibinfo  {journal} {Physical
  Review D}\ }\textbf {\bibinfo {volume} {106}},\ \bibinfo {pages} {053002} (\bibinfo {year} {2022})}\BibitemShut {NoStop}%
\bibitem [{\citenamefont {Gimeno}\ \emph {et~al.}(2023)\citenamefont {Gimeno}, \citenamefont {Isleif}, \citenamefont {Januschek}, \citenamefont {Lindner}, \citenamefont {Meyer}, \citenamefont {Othman}, \citenamefont {Schott}, \citenamefont {Shah},\ and\ \citenamefont {Sohl}}]{gimeno2023tes}%
  \BibitemOpen
  \bibfield  {author} {\bibinfo {author} {\bibfnamefont {J.~A.~R.}\ \bibnamefont {Gimeno}}, \bibinfo {author} {\bibfnamefont {K.-S.}\ \bibnamefont {Isleif}}, \bibinfo {author} {\bibfnamefont {F.}~\bibnamefont {Januschek}}, \bibinfo {author} {\bibfnamefont {A.}~\bibnamefont {Lindner}}, \bibinfo {author} {\bibfnamefont {M.}~\bibnamefont {Meyer}}, \bibinfo {author} {\bibfnamefont {G.}~\bibnamefont {Othman}}, \bibinfo {author} {\bibfnamefont {M.}~\bibnamefont {Schott}}, \bibinfo {author} {\bibfnamefont {R.}~\bibnamefont {Shah}},\ and\ \bibinfo {author} {\bibfnamefont {L.}~\bibnamefont {Sohl}},\ }\href@noop {} {\bibfield  {journal} {\bibinfo  {journal} {Nuclear Instruments and Methods in Physics Research Section A: Accelerators, Spectrometers, Detectors and Associated Equipment}\ }\textbf {\bibinfo {volume} {1046}},\ \bibinfo {pages} {167588} (\bibinfo {year} {2023})}\BibitemShut {NoStop}%
\bibitem [{\citenamefont {Dreyling-Eschweiler}\ \emph {et~al.}(2015)\citenamefont {Dreyling-Eschweiler}, \citenamefont {Bastidon}, \citenamefont {D{\"o}brich}, \citenamefont {Horns}, \citenamefont {Januschek},\ and\ \citenamefont {Lindner}}]{dreyling2015characterization}%
  \BibitemOpen
  \bibfield  {author} {\bibinfo {author} {\bibfnamefont {J.}~\bibnamefont {Dreyling-Eschweiler}}, \bibinfo {author} {\bibfnamefont {N.}~\bibnamefont {Bastidon}}, \bibinfo {author} {\bibfnamefont {B.}~\bibnamefont {D{\"o}brich}}, \bibinfo {author} {\bibfnamefont {D.}~\bibnamefont {Horns}}, \bibinfo {author} {\bibfnamefont {F.}~\bibnamefont {Januschek}},\ and\ \bibinfo {author} {\bibfnamefont {A.}~\bibnamefont {Lindner}},\ }\href {https://doi.org/10.1080/09500340.2015.1021723} {\bibfield  {journal} {\bibinfo  {journal} {Journal of Modern Optics}\ }\textbf {\bibinfo {volume} {62}},\ \bibinfo {pages} {1132} (\bibinfo {year} {2015})}\BibitemShut {NoStop}%
\bibitem [{\citenamefont {Bienfang}\ \emph {et~al.}(2023)\citenamefont {Bienfang}, \citenamefont {Gerrits}, \citenamefont {Kuo}, \citenamefont {Migdall}, \citenamefont {Polyakov},\ and\ \citenamefont {Slattery}}]{dictionary}%
  \BibitemOpen
  \bibfield  {author} {\bibinfo {author} {\bibfnamefont {J.}~\bibnamefont {Bienfang}}, \bibinfo {author} {\bibfnamefont {T.}~\bibnamefont {Gerrits}}, \bibinfo {author} {\bibfnamefont {P.}~\bibnamefont {Kuo}}, \bibinfo {author} {\bibfnamefont {A.}~\bibnamefont {Migdall}}, \bibinfo {author} {\bibfnamefont {S.}~\bibnamefont {Polyakov}},\ and\ \bibinfo {author} {\bibfnamefont {O.}~\bibnamefont {Slattery}},\ }\href@noop {} {\emph {\bibinfo {title} {\href{https://doi.org/10.6028/NIST.IR.8486}{Single-Photon Sources and Detectors Dictionary}}}} (\bibinfo {year} {2023}),\ \bibinfo {note} {{N}IST Internal Report IR 8486, \url{https://nvlpubs.nist.gov/nistpubs/ir/2023/NIST.IR.8486.pdf} [online, accessed 10 Aug 2024]}\BibitemShut {NoStop}%
\bibitem [{\citenamefont {Millar}\ \emph {et~al.}(2017)\citenamefont {Millar}, \citenamefont {Raffelt}, \citenamefont {Redondo},\ and\ \citenamefont {Steffen}}]{millar2017dielectric}%
  \BibitemOpen
  \bibfield  {author} {\bibinfo {author} {\bibfnamefont {A.~J.}\ \bibnamefont {Millar}}, \bibinfo {author} {\bibfnamefont {G.~G.}\ \bibnamefont {Raffelt}}, \bibinfo {author} {\bibfnamefont {J.}~\bibnamefont {Redondo}},\ and\ \bibinfo {author} {\bibfnamefont {F.~D.}\ \bibnamefont {Steffen}},\ }\href {https://doi.org/10.1088/1475-7516/2017/01/061} {\bibfield  {journal} {\bibinfo  {journal} {Journal of Cosmology and Astroparticle Physics}\ }\textbf {\bibinfo {volume} {2017}}\bibinfo  {number} { (01)},\ \bibinfo {pages} {061}}\BibitemShut {NoStop}%
\bibitem [{\citenamefont {Chiles}\ \emph {et~al.}(2022)\citenamefont {Chiles}, \citenamefont {Charaev}, \citenamefont {Lasenby}, \citenamefont {Baryakhtar}, \citenamefont {Huang}, \citenamefont {Roshko}, \citenamefont {Burton}, \citenamefont {Colangelo}, \citenamefont {Van~Tilburg}, \citenamefont {Arvanitaki} \emph {et~al.}}]{chiles2022new}%
  \BibitemOpen
\bibfield  {number} {  }\bibfield  {author} {\bibinfo {author} {\bibfnamefont {J.}~\bibnamefont {Chiles}}, \bibinfo {author} {\bibfnamefont {I.}~\bibnamefont {Charaev}}, \bibinfo {author} {\bibfnamefont {R.}~\bibnamefont {Lasenby}}, \bibinfo {author} {\bibfnamefont {M.}~\bibnamefont {Baryakhtar}}, \bibinfo {author} {\bibfnamefont {J.}~\bibnamefont {Huang}}, \bibinfo {author} {\bibfnamefont {A.}~\bibnamefont {Roshko}}, \bibinfo {author} {\bibfnamefont {G.}~\bibnamefont {Burton}}, \bibinfo {author} {\bibfnamefont {M.}~\bibnamefont {Colangelo}}, \bibinfo {author} {\bibfnamefont {K.}~\bibnamefont {Van~Tilburg}}, \bibinfo {author} {\bibfnamefont {A.}~\bibnamefont {Arvanitaki}}, \emph {et~al.},\ }\href {https://doi.org/https://doi.org/10.1103/PhysRevLett.128.231802} {\bibfield  {journal} {\bibinfo  {journal} {Physical Review Letters}\ }\textbf {\bibinfo {volume} {128}},\ \bibinfo {pages} {231802} (\bibinfo {year} {2022})}\BibitemShut {NoStop}%
\bibitem [{\citenamefont {Manenti}\ \emph {et~al.}(2022)\citenamefont {Manenti}, \citenamefont {Mishra}, \citenamefont {Bruno}, \citenamefont {Roberts}, \citenamefont {Oikonomou}, \citenamefont {Pasricha}, \citenamefont {Sarnoff}, \citenamefont {Weston}, \citenamefont {Arneodo}, \citenamefont {Di~Giovanni} \emph {et~al.}}]{manenti2022search}%
  \BibitemOpen
  \bibfield  {author} {\bibinfo {author} {\bibfnamefont {L.}~\bibnamefont {Manenti}}, \bibinfo {author} {\bibfnamefont {U.}~\bibnamefont {Mishra}}, \bibinfo {author} {\bibfnamefont {G.}~\bibnamefont {Bruno}}, \bibinfo {author} {\bibfnamefont {H.}~\bibnamefont {Roberts}}, \bibinfo {author} {\bibfnamefont {P.}~\bibnamefont {Oikonomou}}, \bibinfo {author} {\bibfnamefont {R.}~\bibnamefont {Pasricha}}, \bibinfo {author} {\bibfnamefont {I.}~\bibnamefont {Sarnoff}}, \bibinfo {author} {\bibfnamefont {J.}~\bibnamefont {Weston}}, \bibinfo {author} {\bibfnamefont {F.}~\bibnamefont {Arneodo}}, \bibinfo {author} {\bibfnamefont {A.}~\bibnamefont {Di~Giovanni}}, \emph {et~al.},\ }\href {https://doi.org/https://doi.org/10.1103/PhysRevD.105.052010} {\bibfield  {journal} {\bibinfo  {journal} {Physical Review D}\ }\textbf {\bibinfo {volume} {105}},\ \bibinfo {pages} {052010} (\bibinfo {year} {2022})}\BibitemShut {NoStop}%
\bibitem [{\citenamefont {Martinis}\ and\ \citenamefont {Clarke}(1985)}]{martinis1985signal}%
  \BibitemOpen
  \bibfield  {author} {\bibinfo {author} {\bibfnamefont {J.~M.}\ \bibnamefont {Martinis}}\ and\ \bibinfo {author} {\bibfnamefont {J.}~\bibnamefont {Clarke}},\ }\href {https://doi.org/https://link.springer.com/article/10.1007/BF00681633} {\bibfield  {journal} {\bibinfo  {journal} {Journal of low temperature physics}\ }\textbf {\bibinfo {volume} {61}},\ \bibinfo {pages} {227} (\bibinfo {year} {1985})}\BibitemShut {NoStop}%
\bibitem [{\citenamefont {Rajteri}\ \emph {et~al.}(2020)\citenamefont {Rajteri}, \citenamefont {Biasotti}, \citenamefont {Faverzani}, \citenamefont {Ferri}, \citenamefont {Filippo}, \citenamefont {Gatti}, \citenamefont {Giachero}, \citenamefont {Monticone}, \citenamefont {Nucciotti},\ and\ \citenamefont {Puiu}}]{rajteri2020tes}%
  \BibitemOpen
  \bibfield  {author} {\bibinfo {author} {\bibfnamefont {M.}~\bibnamefont {Rajteri}}, \bibinfo {author} {\bibfnamefont {M.}~\bibnamefont {Biasotti}}, \bibinfo {author} {\bibfnamefont {M.}~\bibnamefont {Faverzani}}, \bibinfo {author} {\bibfnamefont {E.}~\bibnamefont {Ferri}}, \bibinfo {author} {\bibfnamefont {R.}~\bibnamefont {Filippo}}, \bibinfo {author} {\bibfnamefont {F.}~\bibnamefont {Gatti}}, \bibinfo {author} {\bibfnamefont {A.}~\bibnamefont {Giachero}}, \bibinfo {author} {\bibfnamefont {E.}~\bibnamefont {Monticone}}, \bibinfo {author} {\bibfnamefont {A.}~\bibnamefont {Nucciotti}},\ and\ \bibinfo {author} {\bibfnamefont {A.}~\bibnamefont {Puiu}},\ }\href {https://doi.org/https://link.springer.com/article/10.1007/s10909-019-02271-x} {\bibfield  {journal} {\bibinfo  {journal} {Journal of Low Temperature Physics}\ }\textbf {\bibinfo {volume} {199}},\ \bibinfo {pages} {138} (\bibinfo {year} {2020})}\BibitemShut {NoStop}%
\bibitem [{amu(2024)}]{amumetal}%
  \BibitemOpen
  \href {https://www.amuneal.com/magnetic-shielding/magnetic-shielding-materials-2} {\bibinfo {title} {{Magnetic Shielding Materials - Amuneal: Magnetic Shielding {\&} Custom Fabrication}}} (\bibinfo {year} {2024}),\ \bibinfo {note} {[Online; accessed 22. Apr. 2024]}\BibitemShut {NoStop}%
\bibitem [{\citenamefont {Drung}\ \emph {et~al.}(2007)\citenamefont {Drung}, \citenamefont {Abmann}, \citenamefont {Beyer}, \citenamefont {Kirste}, \citenamefont {Peters}, \citenamefont {Ruede},\ and\ \citenamefont {Schurig}}]{drung2007highly}%
  \BibitemOpen
  \bibfield  {author} {\bibinfo {author} {\bibfnamefont {D.}~\bibnamefont {Drung}}, \bibinfo {author} {\bibfnamefont {C.}~\bibnamefont {Abmann}}, \bibinfo {author} {\bibfnamefont {J.}~\bibnamefont {Beyer}}, \bibinfo {author} {\bibfnamefont {A.}~\bibnamefont {Kirste}}, \bibinfo {author} {\bibfnamefont {M.}~\bibnamefont {Peters}}, \bibinfo {author} {\bibfnamefont {F.}~\bibnamefont {Ruede}},\ and\ \bibinfo {author} {\bibfnamefont {T.}~\bibnamefont {Schurig}},\ }\href@noop {} {\bibfield  {journal} {\bibinfo  {journal} {IEEE Transactions on Applied Superconductivity}\ }\textbf {\bibinfo {volume} {17}},\ \bibinfo {pages} {699} (\bibinfo {year} {2007})}\BibitemShut {NoStop}%
\bibitem [{Mag(2024)}]{Magnicon}%
  \BibitemOpen
  \href@noop {} {\bibinfo {title} {{XXF-1 SQUID Electronics data sheet}}} (\bibinfo {year} {2024}),\ \bibinfo {note} {{M}agnicon website, \url{{http://www.magnicon.com/fileadmin/user_upload/downloads/datasheets/Magnicon_XXF-1.pdf}}, [online, accessed 10 Aug 2024]}\BibitemShut {NoStop}%
\bibitem [{\citenamefont {Wold}\ \emph {et~al.}(1987)\citenamefont {Wold}, \citenamefont {Esbensen},\ and\ \citenamefont {Geladi}}]{wold1987pca}%
  \BibitemOpen
  \bibfield  {author} {\bibinfo {author} {\bibfnamefont {S.}~\bibnamefont {Wold}}, \bibinfo {author} {\bibfnamefont {K.}~\bibnamefont {Esbensen}},\ and\ \bibinfo {author} {\bibfnamefont {P.}~\bibnamefont {Geladi}},\ }\href {https://doi.org/https://doi.org/10.1016/0169-7439(87)80084-9} {\bibfield  {journal} {\bibinfo  {journal} {Chemometrics and Intelligent Laboratory Systems}\ }\textbf {\bibinfo {volume} {2}},\ \bibinfo {pages} {37} (\bibinfo {year} {1987})},\ \bibinfo {note} {proceedings of the Multivariate Statistical Workshop for Geologists and Geochemists}\BibitemShut {NoStop}%
\bibitem [{\citenamefont {Arthur}\ and\ \citenamefont {Vassilvitskii}(2007)}]{Arthur2007kmeans}%
  \BibitemOpen
  \bibfield  {author} {\bibinfo {author} {\bibfnamefont {D.}~\bibnamefont {Arthur}}\ and\ \bibinfo {author} {\bibfnamefont {S.}~\bibnamefont {Vassilvitskii}},\ }in\ \href {https://api.semanticscholar.org/CorpusID:1782131} {\emph {\bibinfo {booktitle} {ACM-SIAM Symposium on Discrete Algorithms}}}\ (\bibinfo {year} {2007})\ \bibinfo {note} {{S}ociety for Industrial and Applied Mathematic, Philadelphia}\BibitemShut {NoStop}%
\bibitem [{\citenamefont {Arneodo}\ \emph {et~al.}(2019)\citenamefont {Arneodo}, \citenamefont {Benabderrahmane}, \citenamefont {Bruno}, \citenamefont {Di~Giovanni}, \citenamefont {Fawwaz}, \citenamefont {Messina},\ and\ \citenamefont {Mussolini}}]{Arneodo_2019}%
  \BibitemOpen
  \bibfield  {author} {\bibinfo {author} {\bibfnamefont {F.}~\bibnamefont {Arneodo}}, \bibinfo {author} {\bibfnamefont {M.}~\bibnamefont {Benabderrahmane}}, \bibinfo {author} {\bibfnamefont {G.}~\bibnamefont {Bruno}}, \bibinfo {author} {\bibfnamefont {A.}~\bibnamefont {Di~Giovanni}}, \bibinfo {author} {\bibfnamefont {O.}~\bibnamefont {Fawwaz}}, \bibinfo {author} {\bibfnamefont {M.}~\bibnamefont {Messina}},\ and\ \bibinfo {author} {\bibfnamefont {C.}~\bibnamefont {Mussolini}},\ }\href {https://doi.org/10.1016/j.nima.2018.09.112} {\bibfield  {journal} {\bibinfo  {journal} {Nuclear Instruments and Methods in Physics Research Section A: Accelerators, Spectrometers, Detectors and Associated Equipment}\ }\textbf {\bibinfo {volume} {936}},\ \bibinfo {pages} {242–243} (\bibinfo {year} {2019})}\BibitemShut {NoStop}%
\bibitem [{\citenamefont {Hagmann}\ \emph {et~al.}(2007)\citenamefont {Hagmann}, \citenamefont {Lange},\ and\ \citenamefont {Wright}}]{Hagmann2007cry}%
  \BibitemOpen
  \bibfield  {author} {\bibinfo {author} {\bibfnamefont {C.}~\bibnamefont {Hagmann}}, \bibinfo {author} {\bibfnamefont {D.}~\bibnamefont {Lange}},\ and\ \bibinfo {author} {\bibfnamefont {D.}~\bibnamefont {Wright}},\ }in\ \href {https://doi.org/10.1109/NSSMIC.2007.4437209} {\emph {\bibinfo {booktitle} {2007 IEEE Nuclear Science Symposium Conference Record}}},\ Vol.~\bibinfo {volume} {2}\ (\bibinfo {year} {2007})\ pp.\ \bibinfo {pages} {1143--1146}\BibitemShut {NoStop}%
\bibitem [{\citenamefont {Feldman}\ and\ \citenamefont {Cousins}(1998)}]{Feldman1998Apr}%
  \BibitemOpen
  \bibfield  {author} {\bibinfo {author} {\bibfnamefont {G.~J.}\ \bibnamefont {Feldman}}\ and\ \bibinfo {author} {\bibfnamefont {R.~D.}\ \bibnamefont {Cousins}},\ }\href {https://doi.org/10.1103/PhysRevD.57.3873} {\bibfield  {journal} {\bibinfo  {journal} {Physical Review D}\ }\textbf {\bibinfo {volume} {57}},\ \bibinfo {pages} {3873} (\bibinfo {year} {1998})}\BibitemShut {NoStop}%
\bibitem [{\citenamefont {Dreyling-Eschweiler}(2014)}]{dreyling2014superconducting}%
  \BibitemOpen
  \bibfield  {author} {\bibinfo {author} {\bibfnamefont {J.}~\bibnamefont {Dreyling-Eschweiler}},\ }\href@noop {} {\emph {\bibinfo {title} {\href{https://s3.cern.ch/inspire-prod-files-8/8fc2b9ec1681bd3b17857ec9052acbbe}{A superconducting microcalorimeter for low-flux detection of near-infrared single photons}}}},\ \bibinfo {type} {Tech. Rep.}\ (\bibinfo  {institution} {Deutsches Elektronen-Synchrotron (DESY)},\ \bibinfo {year} {2014})\BibitemShut {NoStop}%
\bibitem [{git(2024{\natexlab{a}})}]{github}%
  \BibitemOpen
  \href@noop {} {\bibinfo {title} {nyuad astroparticle: tes geant}} (\bibinfo {year} {2024}{\natexlab{a}}),\ \bibinfo {note} {\url{https://github.com/nyuad-astroparticle/tes-geant}}\BibitemShut {NoStop}%
\bibitem [{\citenamefont {Monticone}\ \emph {et~al.}(2021)\citenamefont {Monticone}, \citenamefont {Castellino}, \citenamefont {Rocci},\ and\ \citenamefont {Rajteri}}]{monticone2021ti}%
  \BibitemOpen
  \bibfield  {author} {\bibinfo {author} {\bibfnamefont {E.}~\bibnamefont {Monticone}}, \bibinfo {author} {\bibfnamefont {M.}~\bibnamefont {Castellino}}, \bibinfo {author} {\bibfnamefont {R.}~\bibnamefont {Rocci}},\ and\ \bibinfo {author} {\bibfnamefont {M.}~\bibnamefont {Rajteri}},\ }\href {https://doi.org/10.1109/TASC.2021.3069903} {\bibfield  {journal} {\bibinfo  {journal} {IEEE Transactions on Applied Superconductivity}\ }\textbf {\bibinfo {volume} {31}},\ \bibinfo {pages} {1} (\bibinfo {year} {2021})}\BibitemShut {NoStop}%
\bibitem [{\citenamefont {Virtanen}\ \emph {et~al.}(2020)\citenamefont {Virtanen}, \citenamefont {Gommers}, \citenamefont {Oliphant}, \citenamefont {Haberland}, \citenamefont {Reddy}, \citenamefont {Cournapeau}, \citenamefont {Burovski}, \citenamefont {Peterson}, \citenamefont {Weckesser}, \citenamefont {Bright}, \citenamefont {{van der Walt}}, \citenamefont {Brett}, \citenamefont {Wilson}, \citenamefont {Millman}, \citenamefont {Mayorov}, \citenamefont {Nelson}, \citenamefont {Jones}, \citenamefont {Kern}, \citenamefont {Larson}, \citenamefont {Carey}, \citenamefont {Polat}, \citenamefont {Feng}, \citenamefont {Moore}, \citenamefont {{VanderPlas}}, \citenamefont {Laxalde}, \citenamefont {Perktold}, \citenamefont {Cimrman}, \citenamefont {Henriksen}, \citenamefont {Quintero}, \citenamefont {Harris}, \citenamefont {Archibald}, \citenamefont {Ribeiro}, \citenamefont {Pedregosa}, \citenamefont {{van Mulbregt}},\ and\ \citenamefont {{SciPy 1.0 Contributors}}}]{2020SciPy}%
  \BibitemOpen
  \bibfield  {author} {\bibinfo {author} {\bibfnamefont {P.}~\bibnamefont {Virtanen}}, \bibinfo {author} {\bibfnamefont {R.}~\bibnamefont {Gommers}}, \bibinfo {author} {\bibfnamefont {T.~E.}\ \bibnamefont {Oliphant}}, \bibinfo {author} {\bibfnamefont {M.}~\bibnamefont {Haberland}}, \bibinfo {author} {\bibfnamefont {T.}~\bibnamefont {Reddy}}, \bibinfo {author} {\bibfnamefont {D.}~\bibnamefont {Cournapeau}}, \bibinfo {author} {\bibfnamefont {E.}~\bibnamefont {Burovski}}, \bibinfo {author} {\bibfnamefont {P.}~\bibnamefont {Peterson}}, \bibinfo {author} {\bibfnamefont {W.}~\bibnamefont {Weckesser}}, \bibinfo {author} {\bibfnamefont {J.}~\bibnamefont {Bright}}, \bibinfo {author} {\bibfnamefont {S.~J.}\ \bibnamefont {{van der Walt}}}, \bibinfo {author} {\bibfnamefont {M.}~\bibnamefont {Brett}}, \bibinfo {author} {\bibfnamefont {J.}~\bibnamefont {Wilson}}, \bibinfo {author} {\bibfnamefont {K.~J.}\ \bibnamefont {Millman}}, \bibinfo {author} {\bibfnamefont {N.}~\bibnamefont {Mayorov}}, \bibinfo {author} {\bibfnamefont
  {A.~R.~J.}\ \bibnamefont {Nelson}}, \bibinfo {author} {\bibfnamefont {E.}~\bibnamefont {Jones}}, \bibinfo {author} {\bibfnamefont {R.}~\bibnamefont {Kern}}, \bibinfo {author} {\bibfnamefont {E.}~\bibnamefont {Larson}}, \bibinfo {author} {\bibfnamefont {C.~J.}\ \bibnamefont {Carey}}, \bibinfo {author} {\bibfnamefont {{\.I}.}~\bibnamefont {Polat}}, \bibinfo {author} {\bibfnamefont {Y.}~\bibnamefont {Feng}}, \bibinfo {author} {\bibfnamefont {E.~W.}\ \bibnamefont {Moore}}, \bibinfo {author} {\bibfnamefont {J.}~\bibnamefont {{VanderPlas}}}, \bibinfo {author} {\bibfnamefont {D.}~\bibnamefont {Laxalde}}, \bibinfo {author} {\bibfnamefont {J.}~\bibnamefont {Perktold}}, \bibinfo {author} {\bibfnamefont {R.}~\bibnamefont {Cimrman}}, \bibinfo {author} {\bibfnamefont {I.}~\bibnamefont {Henriksen}}, \bibinfo {author} {\bibfnamefont {E.~A.}\ \bibnamefont {Quintero}}, \bibinfo {author} {\bibfnamefont {C.~R.}\ \bibnamefont {Harris}}, \bibinfo {author} {\bibfnamefont {A.~M.}\ \bibnamefont {Archibald}}, \bibinfo {author}
  {\bibfnamefont {A.~H.}\ \bibnamefont {Ribeiro}}, \bibinfo {author} {\bibfnamefont {F.}~\bibnamefont {Pedregosa}}, \bibinfo {author} {\bibfnamefont {P.}~\bibnamefont {{van Mulbregt}}},\ and\ \bibinfo {author} {\bibnamefont {{SciPy 1.0 Contributors}}},\ }\href {https://doi.org/10.1038/s41592-019-0686-2} {\bibfield  {journal} {\bibinfo  {journal} {Nature Methods}\ }\textbf {\bibinfo {volume} {17}},\ \bibinfo {pages} {261} (\bibinfo {year} {2020})}\BibitemShut {NoStop}%
\bibitem [{git(2024{\natexlab{b}})}]{github2}%
  \BibitemOpen
  \href@noop {} {\bibinfo {title} {nyuad astroparticle: tes analysis}} (\bibinfo {year} {2024}{\natexlab{b}}),\ \bibinfo {note} {\url{https://github.com/nyuad-astroparticle/TES-analysis}}\BibitemShut {NoStop}%
\bibitem [{\citenamefont {Agostinelli}\ \emph {et~al.}(2003)\citenamefont {Agostinelli}, \citenamefont {Allison}, \citenamefont {Amako}, \citenamefont {Apostolakis}, \citenamefont {Araujo}, \citenamefont {Arce}, \citenamefont {Asai}, \citenamefont {Axen}, \citenamefont {Banerjee}, \citenamefont {Barrand} \emph {et~al.}}]{agostinelli2003geant4}%
  \BibitemOpen
  \bibfield  {author} {\bibinfo {author} {\bibfnamefont {S.}~\bibnamefont {Agostinelli}}, \bibinfo {author} {\bibfnamefont {J.}~\bibnamefont {Allison}}, \bibinfo {author} {\bibfnamefont {K.~a.}\ \bibnamefont {Amako}}, \bibinfo {author} {\bibfnamefont {J.}~\bibnamefont {Apostolakis}}, \bibinfo {author} {\bibfnamefont {H.}~\bibnamefont {Araujo}}, \bibinfo {author} {\bibfnamefont {P.}~\bibnamefont {Arce}}, \bibinfo {author} {\bibfnamefont {M.}~\bibnamefont {Asai}}, \bibinfo {author} {\bibfnamefont {D.}~\bibnamefont {Axen}}, \bibinfo {author} {\bibfnamefont {S.}~\bibnamefont {Banerjee}}, \bibinfo {author} {\bibfnamefont {G.}~\bibnamefont {Barrand}}, \emph {et~al.},\ }\href {https://doi.org/https://doi.org/10.1016/S0168-9002(03)01368-8} {\bibfield  {journal} {\bibinfo  {journal} {Nuclear instruments and methods in physics research section A: Accelerators, Spectrometers, Detectors and Associated Equipment}\ }\textbf {\bibinfo {volume} {506}},\ \bibinfo {pages} {250} (\bibinfo {year} {2003})}\BibitemShut {NoStop}%
\end{thebibliography}%

\end{document}